\documentclass[journal]{IEEEtran}
\IEEEoverridecommandlockouts
\usepackage{amsmath, amsfonts, amssymb}
\usepackage{algorithmic}
\usepackage[dvipsnames]{xcolor}
\usepackage{algorithm}
\usepackage{array}
\usepackage{textcomp}
\usepackage{stfloats}
\usepackage{url}
\usepackage[hidelinks]{hyperref}
\usepackage{verbatim}
\usepackage{graphicx, color, xcolor}
\usepackage{cite}
\usepackage{subcaption}
\usepackage[thinc]{esdiff}
\usepackage{lipsum}
\usepackage{enumerate} 
\usepackage{multirow}
\usepackage{pifont}

\usepackage{tabularx}
\usepackage{booktabs} 

\begin{document}

\title{Site Geometry and Calibration Uncertainties in Digital Twin-enabled Channel Estimation}
\author{Lorenzo Del Moro, Francesco Linsalata, Umberto Spagnolini and Maurizio Magarini \\
{\textit{DEIB, Politecnico di Milano, Milan, Italy}}
{Email: \{name.surname\}@polimi.it},
\thanks{The work was partially supported by the European Union under the Italian National Recovery and Resilience Plan (NRRP) of NextGenerationEU, partnership on “Telecommunications of the Future” (PE00000001 - program “RESTART”, Structural Project 6GWINET). This work is within the Joint Lab between Politecnico di Milano and Huawei.}}



\maketitle

\begin{abstract}
Fast ray tracing (RT) has stimulated the Digital Twin (DT) as an emerging technology for environment-aware communications. 
Since wireless propagation is governed by the interaction between site geometry and electromagnetic (EM) properties of the environment, DT-based approaches can provide site-specific prior information for channel estimation. In this work, we investigate the robustness of DT to aid the channel estimation, where multipath features extracted via RT are used to construct the low-rank (LR) eigenstructure of the channel covariance matrix. This LR structure is used in channel estimation. However, the digital representation of propagation model is inaccurate and thus it affects the LR. We explicitly analyze these model mismatches that arise from user positioning errors, which translate into geometric inconsistencies in the site representation, and EM material calibration errors. We derive a first-order perturbative model that separates geometric perturbations, affecting angles and delays, from EM perturbations, affecting path gains. Based on this perturbed model, we provide a normalized mean-square error (NMSE) analysis that reveals a fundamental difference between geometric and EM perturbations. In particular, we show that LR estimation is inherently robust to EM calibration perturbations, while positioning errors, dominate performance degradation by altering the channel eigenstructure. Numerical results confirm that, in urban, suburban and rural scenarios, positioning errors are the primary limiting factor, whereas EM calibration errors have a 
comparatively limited impact. Despite these mismatches, DT-empowered estimators provide up to $10\:\text{dB}$ NMSE improvement, over baseline methods, in the urban low signal-to-noise ratio (SNR) settings, while achieving performance comparable to baseline estimators at high SNR for moderate ($<1\:\text{m}$) positioning errors.
\end{abstract}

\begin{IEEEkeywords}
Digital Twin, Ray tracing, Channel estimation
\end{IEEEkeywords}

\section{Introduction}
\label{Introduction}

Next-generation networks are expected to support new applications and use cases, imposing stringent requirements in terms of data rates, reliability, and latency. To meet these challenging requirements, higher carrier frequencies, e.g., upper mid-band and millimeter-wave (mmWave), are expected to support large bandwidths despite severe propagation path loss~\cite{Jiang_6G}. Multiple-input multiple-output (MIMO) is regarded as a key technology for enabling the shift to these frequency bands. However, wideband MIMO channels are characterized by a large number of parameters, making accurate channel estimation increasingly challenging~\cite{Brighente_RR}. Low-rank (LR) and minimum mean square error (MMSE) estimators have been proposed to address this issue by exploiting the eigenstructure of the channel covariance matrix, which typically exhibits LR properties due to the sparse structure of wideband MIMO channels in the spatio-temporal (ST) domain~\cite{Xie_sparsity}.

From a physical standpoint, this eigenstructure is determined by the underlying propagation geometry and by the electromagnetic interactions with the environment, which jointly define the dominant multipath components.
In particular, the site geometry, together with the Base Station (BS) and User Equipment (UE) positions, determines the propagation paths, so that inaccuracies in UE positioning directly translate into geometric inconsistencies in the effective propagation scenario seen by the BS.
Unfortunately, acquiring reliable covariance information remains challenging in practice, as its dimension scales with the channel dimensionality and channel statistics~\cite{Mizmizi_V2X}. These challenges have motivated several complementary research directions, including learning-assisted channel estimation~\cite{Li_THz_ch_est,Li_THz_ch_est_2}, advanced wave-domain processing through stacked intelligent metasurfaces (SIMs)~\cite{Li_SIM,Wei_SIM}, and environment-aware communications that exploit prior knowledge of the propagation environment~\cite{Zeng_env_aware}.

Recent research on environment-aware communications has increasingly leveraged the Digital Twin (DT), a virtual representation that mirrors the components and interactions of the physical system~\cite{Lin_DT}. In the radio access network (RAN) context, DT systems enable the synthesis of prior channel information by emulating the electromagnetic propagation environment, with ray tracing (RT) acting as a key enabling tool~\cite{cazzella2025high, FangDT}. By exploiting high-fidelity digital models of the scenario, obtained for instance via multimodal sensing or geospatial databases~\cite{Kamioka_Lidar,cazzella2025high}, RT simulations extract relevant parameters of the dominant multipath components for given BS and UE positions, accounting for reflection, diffraction, and diffuse scattering. Compared to purely data-driven approaches, RT leverages a priori knowledge grounded in the physics of electromagnetic propagation, offering the potential to accurately capture angles-of-arrival (AoAs), angles-of-departure (AoDs), propagation delays, and path gains~\cite{Ruah_RT}. These parameters fully determine the channel covariance matrix and its eigenstructure, which represent the key quantities exploited by LR and MMSE estimators. As a result, any mismatch in the digital representation of the environment propagates to the estimated eigenstructure, potentially degrading the performance of DT-empowered channel estimation.

\textbf{Related works}
Several recent works have investigated the use of the DT in the RAN, focusing either on improving the modeling fidelity and computational efficiency of RT simulations~\cite{SionnaRT2023,cazzella2025high,Zhu_RT} or on architectural aspects such as DT orchestration, real-time synchronization, and predictive optimization within the RAN~\cite{Wu_DT,Lin_DT}. More recently, DTs have also been explored for tasks such as localization, hybrid precoding/combining, and CSI feedback~\cite{Luo_DT_Asilomar,Morais_DT,Luo_DT}.

DT has also been increasingly adopted to support channel estimation using site-specific propagation information. For example, in~\cite{alikhani_DT}, DT-derived subspaces are iteratively calibrated through reinforcement learning (RL) to enable zone-specific subspace-based channel estimation, thereby reducing feedback overhead. In our previous works~\cite{DelMoro_DT,DelMoro_DT_2}, we exploited the ST eigenstructure reconstructed from the DT as a prior for uplink LR channel estimation and investigated the impact of the limited propagation knowledge arising from the computational constraints of real-time RT.

In parallel, several works have addressed the problem of calibrating RT-based DTs to better match real channel observations. In particular,~\cite{Ruah_RT} shows that even small discrepancies in the site geometry can induce significant phase errors in the RT-generated multipath components, and proposes a channel-response-based calibration method to compensate for such effects. More generally, RT calibration techniques aim at refining the EM parameters of the digital model using channel measurements~\cite{Hoydis_RT,Calibration_1}. These approaches improve the fidelity of RT predictions by calibrating the simulated channel responses, but do not analyze how residual mismatches affect the channel eigenstructure and, in turn, channel estimation performance.

Hence, existing DT-assisted communication frameworks either assume an ideal or calibrated digital representation of the propagation environment or focus on improving RT fidelity, without analytically characterizing how residual DT inaccuracies propagate to the channel covariance eigenstructure and affect LR and MMSE channel estimation performance. Table~I summarizes the main differences between representative existing works and the proposed framework.

Although high-resolution 3D models can be obtained~\cite{SionnaRT2023}, residual mismatches may still arise due to imperfect UE positioning or inaccuracies in the EM characterization of the environment. In particular, UE positioning errors introduce geometric inconsistencies in the reconstructed propagation scenario, whereas EM calibration errors primarily perturb the multipath gains while preserving the propagation geometry under the small-error assumption adopted in this work. Accordingly, these two error sources induce fundamentally different perturbations of the covariance eigenstructure, resulting in a distinct impact on LR and MMSE channel estimation.

Assessing these effects experimentally is challenging because geometry, positioning, material properties, and RT modeling errors are inherently coupled in practical deployments. Consequently, measurement-based evaluations do not easily allow isolating the individual contribution of geometric and EM mismatches or controlling their magnitude. This motivates the use of controlled high-fidelity RT-based simulations, which enable a systematic and interpretable assessment of the individual impact of geometric and EM mismatches while isolating RT approximation errors from DT mismatch effects.

To address this gap, this work proposes a model-based framework that links DT inaccuracies to perturbations of the channel covariance eigenstructure. Specifically, we adopt a first-order perturbation approach that separates geometric mismatches induced by UE positioning errors from EM mismatches affecting path gains, and analyze their impact on DT-empowered LR and MMSE estimation.

\begin{table*}[t]
\centering
\caption{Comparison between representative existing works and the proposed framework.}
\renewcommand{\arraystretch}{1.2}
\small
\begin{tabular}{p{3.6cm}|p{6.6cm}|p{3.0cm}|c}
\hline
\textbf{Reference} &
\textbf{Main contribution} &
\textbf{Channel estimation} &
\textbf{Perturbation analysis} \\
\hline
\cite{SionnaRT2023,cazzella2025high,Wu_DT,Lin_DT,Zhu_RT} &
RT-based DT construction, modeling fidelity, orchestration, and RAN optimization &
\ding{55} &
\ding{55} \\
\hline
\cite{Hoydis_RT,Ruah_RT,Calibration_1} &
Calibration of RT-based DTs against real-world measurements &
\ding{55} &
\ding{55} \\
\hline
\cite{Luo_DT_Asilomar,Morais_DT,Luo_DT} &
DT-assisted localization, hybrid precoding, and CSI feedback &
\ding{55} &
\ding{55} \\
\hline
\cite{alikhani_DT} &
DT-derived subspaces with RL for channel estimation &
Subspace-based &
\ding{55} \\
\hline
\cite{DelMoro_DT,DelMoro_DT_2} &
DT-derived channel priors for channel estimation &
LR &
\ding{55} \\
\hline
\textbf{This work} &
Perturbation analysis of DT inaccuracies for DT-empowered channel estimation &
LR/MMSE &
\ding{51} \\
\hline
\end{tabular}
\label{tab:comparison}
\end{table*}

\textbf{Paper Contributions} The main contributions of this paper are the following:
\begin{itemize}

\item
We propose a DT-empowered channel estimation framework that integrates high-fidelity 3D models and relative RT. The framework extracts dominant multipath parameters to reconstruct the channel eigenstructure for a given BS and UE position, which serves for DT-empowered LR and MMSE channel estimation.

\item
We develop perturbation models for the multipath parameters under small-error regimes. In particular, we derive (i) AoA/AoD and delay perturbations due to UE positioning errors and (ii) path gain perturbations due to calibration errors in the EM modeling of reflections.

\item
We provide an NMSE analysis to gain insights into the impact of mismatches on LR and MMSE channel estimation in MIMO-OFDM systems, showing that LR estimators are loosely insensitive to EM calibration-induced gain perturbations under preserved channel subspace conditions, while positioning errors induce subspace distortions that directly impact estimation performance.

\item
We validate the proposed framework via RT simulations, in high-fidelity urban, suburban, and rural scenarios. Results show that the proposed DT-empowered estimators remain robust to realistic positioning and EM calibration errors, achieving up to $10\:\mathrm{dB}$ NMSE improvement over baseline approaches in the low SNR region.

\end{itemize}

\textbf{Paper Organization}. 
The remainder of the paper is organized as follows. Section~\ref{System Model and Channel Model} introduces the uplink MIMO-OFDM model and the corresponding ST channel representation. Section~\ref{Digital Twin Framework} describes the DT-empowered channel estimation framework and the LR and MMSE estimators. Section~\ref{sec:DT_perturbations_metrics_nmse} presents the perturbation models and the NMSE analysis, while Section~\ref{Numerical results} reports simulation results. Finally, Section~\ref{sec:conclusion} concludes the paper.

\textbf{Notation}. 
$(\cdot)^{\mathsf{*}}$, $(\cdot)^{\mathsf{T}}$, and $(\cdot)^{\mathsf{H}}$ denote complex conjugate, transpose, and Hermitian transpose, respectively, and $(\cdot)^{\dagger}$ the Moore-Penrose pseudoinverse. The Kronecker product is denoted by $\otimes$, $\mathrm{diag}(\cdot)$ forms a diagonal matrix, and $\mathrm{vec}(\cdot)$ denotes vectorization. $\mathrm{Tr}\{\cdot\}$, $\mathrm{rank}\{\cdot\}$, and $\mathrm{span}\{\cdot\}$ denote trace, rank, and column space, respectively, while $\mathbb{E}\{\cdot\}$ denotes expectation and $\mathbf{I}_N$ the identity matrix. The norm and inner product are $\|\cdot\|$ and $\langle \mathbf{a},\mathbf{b}\rangle=\mathbf{a}^{\mathsf{H}}\mathbf{b}$. $\mathcal{N}(\cdot,\cdot)$ and $\mathcal{N}_{\mathbb{C}}(\cdot,\cdot)$ denote real and complex Gaussian distributions, respectively, $\delta(\cdot)$ is the Kronecker delta, and $\Re\{\cdot\}$ denotes the real part.

\section{System and Channel Model}
\label{System Model and Channel Model}

In this section, we first describe the uplink pilot transmission in an $N_\mathrm{r} \times N_\mathrm{t}$ MIMO-OFDM system and then provide the ST channel model. The BS and the UE are equipped with $N_{\mathrm{r}}$ and $N_{\mathrm{t}}$ antenna elements, respectively. The duration of an OFDM symbol is given by $T = (N_{\mathrm{sc}} + N_{\mathrm{cp}}) T_{\mathrm{s}}$, where $T_{\mathrm{s}} = 1/B$ represents the sampling interval for a system bandwidth $B$, $N_{\mathrm{cp}}$ is the length of the cyclic prefix (CP), and $N_{\mathrm{sc}}$ denotes the number of subcarriers.

\subsection{Uplink Pilot Transmission}
\label{Uplink Pilot Transmission and ML}

We assume that channel estimation is performed using $N_\mathrm{t}$ mutually orthogonal pilot sequences, each transmitted from a different UE antenna to the BS. These sequences, known at the BS, consist of $N_\mathrm{p}$ pilot samples, scattered on a frequency grid, corresponding to predetermined subcarriers within an OFDM training symbol. We consider the transmission of $M$ OFDM training symbols, indexed by $m=1,\dots, M$. Let $\mathbf{x}^\ell \in \mathbb{C}^{N_\mathrm{p}}$ denote the frequency-domain baseband pilot sequence transmitted from the $\ell$th antenna ($\ell = 1,\dots,N_{\mathrm{t}}$) in each training symbol, such that
$(\mathbf{x}^\ell)^\mathsf{H}\mathbf{x}^{\ell'} = P_{\mathrm{T}} N_\mathrm{p}\,\delta(\ell-\ell')$,
where $P_{\mathrm{T}}$ denotes the average power per pilot subcarrier. We assume a CP $N_{\mathrm{cp}}\geq W-1$, where $W$ is the maximum temporal support of the channel impulse response (CIR). Accordingly, the vector $\mathbf{y}^q(m)\in\mathbb{C}^{N_{\mathrm{p}}}$ of complex baseband received pilot samples in the frequency-domain, at the $q$th antenna ($q = 1,\dots, N_{\mathrm{r}}$) and for the $m$th training symbol, can be written as
\begin{equation}
\label{eq:rx_sum}
\mathbf{y}^q(m)=\sum_{\ell=1}^{N_{\mathrm{t}}}\mathrm{diag}\big(\mathbf{x}^{\ell}\big){\mathbf{h}}_F^{\ell,q}(m)+\mathbf{w}^q(m),
\end{equation}
where ${\mathbf{h}}_F^{\ell,q}(m)\in\mathbb{C}^{N_\mathrm{p}}$ is the equivalent baseband discrete-frequency channel response for the $(\ell,q)$th link, for the $N_\mathrm{p}$ predetermined subcarriers. This can be written as ${\mathbf{h}}_F^{\ell,q}(m) =\widetilde{\mathbf{F}}\,\mathbf{h}^{\ell,q}(m)$, where $\mathbf{h}^{\ell,q}(m)\in\mathbb{C}^{W}$ denotes the discrete-time CIR for the $(\ell,q)$th link and $\widetilde{\mathbf{F}}$ is the partial Discrete Fourier Transform, with entries 
\begin{equation}
    [\widetilde{\mathbf{F}}]_{k,w} = \frac{1}{\sqrt{N_\mathrm{sc}}}\exp\left(-j2\pi k w / N_\mathrm{sc}\right),
\end{equation}
where $k\in\{k_0,\dots,k_{N_\mathrm{p}-1}\}$ denotes the pilot subcarrier index and $w=0,\dots,W-1$. Moreover, $\mathbf{w}^q(m) \sim \mathcal{N}_{\mathbb{C}}(\mathbf{0},\sigma^2_w \mathbf{I}_{N_\mathrm{p}})$ is additive white Gaussian noise (AWGN) with variance $\sigma^2_w$. By stacking the CIRs as
$\mathbf{h}^q(m)=\big[\mathbf{h}^{1,q}(m)^{\mathsf{T}},\dots,\mathbf{h}^{N_{\mathrm{t}},q}(m)^{\mathsf{T}}\big]^{\mathsf{T}}\in\mathbb{C}^{WN_{\mathrm{t}}}$,
\eqref{eq:rx_sum} can be compactly rewritten as 
\begin{equation}
    \label{eq:rx_compact}
\mathbf{y}^q(m)=\mathbf{B}\,\mathbf{h}^q(m)+\mathbf{w}^q(m),
\end{equation}
with $\mathbf{B}=\big[\mathrm{diag}\big(\mathbf{x}^{1}\big)\widetilde{\mathbf{F}},\dots,\mathrm{diag}\big(\mathbf{x}^{N_{\mathrm{t}}}\big)\widetilde{\mathbf{F}}\big]\in\mathbb{C}^{N_{\mathrm{p}}\times WN_{\mathrm{t}}}$.
Let $\mathbf{Y}(m)=\big[\mathbf{y}^1(m),\dots,\mathbf{y}^{N_\mathrm{r}}(m)\big]\in \mathbb{C}^{N_\mathrm{p}\times N_\mathrm{r}}$ be the collection of pilot samples received on the $N_\mathrm{r}$ receiving antennas. Stacking the columns of $\mathbf{Y}(m)$ yields
\begin{equation}
    \label{eq: Received pilot samples (2)}
    \begin{split}
    \mathbf{y}(m) = \mathrm{vec}\left(\mathbf{Y}(m)\right) =\mathbf{\Psi}\mathbf{h}(m) + \mathbf{w}(m),\\          
    \end{split}
\end{equation}
where $\mathbf{h}(m)=\big[\mathbf{h}^1(m)^\mathsf{T},\dots,\mathbf{h}^{N_\mathrm{r}}(m)^\mathsf{T}\big]^\mathsf{T}\in \mathbb{C}^{L}$, with $L=WN_\mathrm{t}N_\mathrm{r}$, is the ST MIMO channel during the $m$th training symbol, while the transmitted complex samples for piloting become $\mathbf{\Psi}=\mathbf{I}_{N_\mathrm{r}}\otimes\mathbf{B}\in\mathbb{C}^{N_\mathrm{r}N_\mathrm{p} \times L}$, and $\mathbf{w}(m)\sim\mathcal{N}_{\mathbb{C}}(\mathbf{0}_{N_\mathrm{p}N_\mathrm{r}},\sigma^2_w\mathbf{I}_{N_\mathrm{p}N_\mathrm{r}})$ is AWGN. A necessary condition for identifiability in channel estimation is $N_{\mathrm{p}} \geq W N_{\mathrm{t}}$. Furthermore, optimal pilot design for maximum-likelihood (ML) channel estimation results in equipowered, equispaced, and phase shift orthogonal pilot samples, such that $\mathbf{\Psi}^\mathsf{H}\mathbf{\Psi} =P_{\mathrm{T}} N_\mathrm{p}/N_\mathrm{sc}\mathbf{I}_{L}=\nu\mathbf{I}_{L}$. Under Gaussian settings, the ML estimate is~\cite{Gao_ML}
\begin{equation}
\label{eq: ML}
\widehat{\mathbf{h}}_\mathrm{ML}(m) = \mathbf{\Psi}^\mathsf{\dagger}\mathbf{y}(m),
\end{equation}
where $ \mathbf{\Psi}^\mathsf{\dagger}\in \mathbb{C}^{L \times N_\mathrm{r}N_\mathrm{p}}$ is the pseudo-inverse of the transmitted pilot matrix in~\eqref{eq: Received pilot samples (2)}.

\subsection{ST MIMO Channel Model}
\label{MIMO-OFDM channel model}

The wireless channel is characterized using a geometric propagation model, where the channel is represented as the superposition of $P$ discrete multipath components, each associated with a planar wave and characterized by its AoA, AoD, propagation delay, and path gain. For the $p$th path ($p=1,\dots,P$), let $\boldsymbol{\theta}_p = [\theta^{\mathrm{az}}_p,\theta^{\mathrm{el}}_p] \in [-\pi,\pi) \times [-\pi/2,\pi/2)$ denote the azimuth and elevation AoA, and $\boldsymbol{\gamma}_p = [\gamma^{\mathrm{az}}_p,\gamma^{\mathrm{el}}_p] \in [-\pi,\pi) \times [-\pi/2,\pi/2)$ the azimuth and elevation AoD, both expressed in a global coordinate system (GCS). Similarly, let $\tau_p \in \mathbb{R}_{\geq 0}$ denote the propagation delay of the $p$th path and $\Omega_p \in \mathbb{R}_{\geq 0}$ the corresponding path gain. $\boldsymbol{\xi}_p = [\boldsymbol{\theta}_p,\boldsymbol{\gamma}_p,\tau_p]^\mathsf{T}$ collects all the ST features of the $p$th propagation path, i.e., AoA, AoD, and delay. Let $\boldsymbol{\xi} = [\boldsymbol{\xi}_1^\mathsf{T},\dots,\boldsymbol{\xi}_P^\mathsf{T}]^\mathsf{T}$ and $\boldsymbol{\Omega} = [\Omega_1,\dots,\Omega_P]^\mathsf{T}$ denote the vectors collecting the ST features and gains of the $P$ multipath components, respectively. The channel parameters $\{\boldsymbol{\xi},\boldsymbol{\Omega}\}$ are assumed constant over the $M$ training symbols, consistently with the ST resolution of the OFDM-MIMO system~\cite{Brighente_RR}.

For a BS located in the far-field region of the UE, the $N \times 1$ array response vectors at the BS, can be modeled through a planar wavefront model~\cite{Bjornsson_WC}:
\begin{equation}
    \label{array_response_BS}
    \mathbf{a}_\mathrm{BS}(\boldsymbol{\theta}_p)=
    \begin{bmatrix}e^{j\frac{2\pi}{\lambda}{(\mathbf{u}^\mathrm{BS}_1})^\mathsf{T}\mathbf{v}^\mathrm{BS}_p},\dots,e^{j\frac{2\pi}{\lambda}{(\mathbf{u}^\mathrm{BS}_{N_\mathrm{r}}})^\mathsf{T}\mathbf{v}^\mathrm{BS}_p}
    \end{bmatrix}^\mathsf{T},
\end{equation}
where $\lambda$ is the carrier wavelength, $\mathbf{u}^\mathrm{BS}_q$ is the 3D position of the $q$th array element in the GCS and $\mathbf{v}^\mathrm{BS}_p=\begin{bmatrix}\cos(\theta^{\mathrm{az}}_p)\cos(\theta^{\mathrm{el}}_p),\sin(\theta^{\mathrm{az}}_p)\cos(\theta^{\mathrm{el}}_p),\sin(\theta^{\mathrm{el}}_p)\end{bmatrix}^{\mathsf{T}}$ denotes the unit-norm AoA vector pointing at the BS array. Similarly, the array response at the UE, for the $p$th path, is 
\begin{equation}
    \label{array_response_UE}
    \mathbf{a}_\mathrm{UE}(\boldsymbol{\gamma}_p)=
    \begin{bmatrix}e^{j\frac{2\pi}{\lambda}{(\mathbf{u}^\mathrm{UE}_1})^\mathsf{T}\mathbf{v}^\mathrm{UE}_p},...,e^{j\frac{2\pi}{\lambda}{(\mathbf{u}^\mathrm{UE}_{N_\mathrm{t}}})^\mathsf{T}\mathbf{v}^\mathrm{UE}_p}
    \end{bmatrix}^\mathsf{T},
\end{equation}
where $\mathbf{u}^\mathrm{UE}_\ell $ is the 3D position of the $\ell$th array element in the GCS, and 
$\mathbf{v}^\mathrm{UE}_p = [\cos(\gamma^{\mathrm{az}}_p)\cos(\gamma^{\mathrm{el}}_p), 
\sin(\gamma^{\mathrm{az}}_p)\cos(\gamma^{\mathrm{el}}_p), 
\sin(\gamma^{\mathrm{el}}_p)]^\mathsf{T}$ 
is the unit-norm AoD vector pointing away from the UE array. We define as $\mathbf{g}(\tau_p)=\begin{bmatrix}g(-\tau_p),\dots,g((W-1)T_s-\tau_p)\end{bmatrix}^\mathsf{T}\in\mathbb{R}^W$ the discrete-time CIR given by the cascade of the BS and UE pulse shaping filters, for the $p$th path. Accordingly, we model the equivalent discrete-time ST MIMO channel $\mathbf{h}(m)$ in~\eqref{eq: Received pilot samples (2)} between the UE and the BS during the $m$th training symbol as
\begin{equation}
\label{eq: overall ST MIMO channel}
\mathbf{h}(m)=\sum_{p=1}^{P}\sqrt{\Omega_p}\Big(
\mathbf{a}_\mathrm{BS}(\boldsymbol{\theta}_p)\!\otimes\!
\mathbf{a}_\mathrm{UE}(\boldsymbol{\gamma}_p)\!\otimes\!
\mathbf{g}(\tau_p)
\Big)e^{j\phi_p(m)}
\end{equation}
where $e^{j\phi_p(m)}$ denotes the phase term associated with the $p$th propagation path, and $\boldsymbol{\zeta}(m) = \big[e^{j\phi_1(m)},\dots,e^{j\phi_P(m)}\big]^\mathsf{T}$ collects the $P$ phase terms, where $\{\phi_p(m)\}_{p=1}^P$ are i.i.d. and $\sim \mathcal{U}[-\pi,\pi)$~\cite{Ruah_RT}. In~\eqref{eq: overall ST MIMO channel}, the channel parameters $\{\boldsymbol{\xi},\boldsymbol{\Omega}\}$ capture the slow time-varying structure of the channel that remain invariant over the $M$ training symbols, whereas $\boldsymbol{\zeta}(m)$ accounts for the fast time-varying fluctuations of the multipath components across training symbols. These are mutually uncorrelated, i.e.,
$\mathbb{E}\{\boldsymbol{\zeta}(m)\boldsymbol{\zeta}^{\mathsf{H}}(m)\}=\mathbf{I}_{P}$ and independent across symbols, i.e.,
$\mathbb{E}\{\boldsymbol{\zeta}(m)\boldsymbol{\zeta}^{\mathsf{H}}(m+l)\}=\mathbf{I}_{P}\delta(l)$. Accordingly, the ST channel covariance matrix reads
\begin{equation}
\label{eq: ST covariance matrix(2)}
\mathbf{R}(\boldsymbol{\xi},\boldsymbol{\Omega})
=
\mathbf{T}(\boldsymbol{\xi})\,\mathrm{diag}(\boldsymbol{\Omega})\,\mathbf{T}(\boldsymbol{\xi})^\mathsf{H},
\end{equation}
where $\mathbf{T}(\boldsymbol{\xi})=[\mathbf{t}(\boldsymbol{\xi}_1),\dots,\mathbf{t}(\boldsymbol{\xi}_P)]\in\mathbb{C}^{L\times P}$ collects the ST responses of the $P$ propagation paths, with $\mathbf{t}(\boldsymbol{\xi}_p)=\mathbf{a}_\mathrm{BS}(\boldsymbol{\theta}_p)\otimes
\mathbf{a}_\mathrm{UE}(\boldsymbol{\gamma}_p)\otimes\mathbf{g}(\tau_p)\in\mathbb{C}^{L}$. $\mathbf{T}(\boldsymbol{\xi})$ depends only on the ST features, whereas $\mathrm{diag}(\boldsymbol{\Omega})$ determines the power distribution across the propagation paths. The eigenvalue decomposition (EVD) of $\mathbf{R}(\boldsymbol{\xi},\boldsymbol{\Omega})$ is $\mathbf{R}(\boldsymbol{\xi},\boldsymbol{\Omega})=\mathbf{U}\mathbf{\Lambda}\mathbf{U}^\mathsf{H}$, where $\mathbf{\Lambda}=\mathrm{diag}(\lambda_1,\dots,\lambda_r)$ contains the $r\leq P$ non-zero eigenvalues and $\mathbf{U}=[\mathbf{u}_1,\dots,\mathbf{u}_r]\in\mathbb{C}^{L\times r}$ collects the corresponding orthonormal eigenvectors. The pair $\{\mathbf{U},\mathbf{\Lambda}\}$ defines the covariance eigenstructure, whose geometric and spectral components are given by the channel subspace $\mathcal{S}\equiv\mathrm{span}\{\mathbf{U}\}\equiv\mathrm{span}\{\mathbf{R}(\boldsymbol{\xi},\boldsymbol{\Omega})\}\equiv\mathrm{span}\{\mathbf{T}(\boldsymbol{\xi})\}$ and the eigenvalue profile $\mathbf{\Lambda}$, respectively. The subspace dimension is $r=\mathrm{rank}\{\mathbf{R}(\boldsymbol{\xi},\boldsymbol{\Omega})\}=\mathrm{rank}\{\mathbf{T}(\boldsymbol{\xi})\}$, corresponding to the number of resolvable ST propagation modes supported by the system resolution. Since wideband MIMO channels are typically characterized by a sparse set of impulse responses in the ST domain, particularly at mmWave frequencies~\cite{Xie_sparsity}, the LR eigenstructure is exploited by conventional LR and MMSE channel estimators. This provides the basis for the DT framework introduced in the next section, where the channel parameters $\{\boldsymbol{\xi},\boldsymbol{\Omega}\}$ are extracted by integrating a digital model of the environment and a RT simulation, to construct a prior estimate of the channel covariance and corresponding eigenstructure.

\section{Digital Twin-based Channel Estimation}
\label{Digital Twin Framework}

This section presents the DT-empowered channel estimation framework used throughout the paper. We analyze how inaccuracies in the DT propagate to the channel eigenstructure and, in turn, affect channel estimation performance. First, we describe the DT deployed at the BS and its role in extracting relevant channel parameters. Next, we detail the overall DT framework. Finally, we introduce the DT-empowered LR and MMSE channel estimation methods, which will later be studied under geometric and EM perturbations.

\subsection{Channel Estimation Framework}
\label{DT_subsec_generation}

\begin{figure}[!t]
\centering
\includegraphics[width=0.5\textwidth]{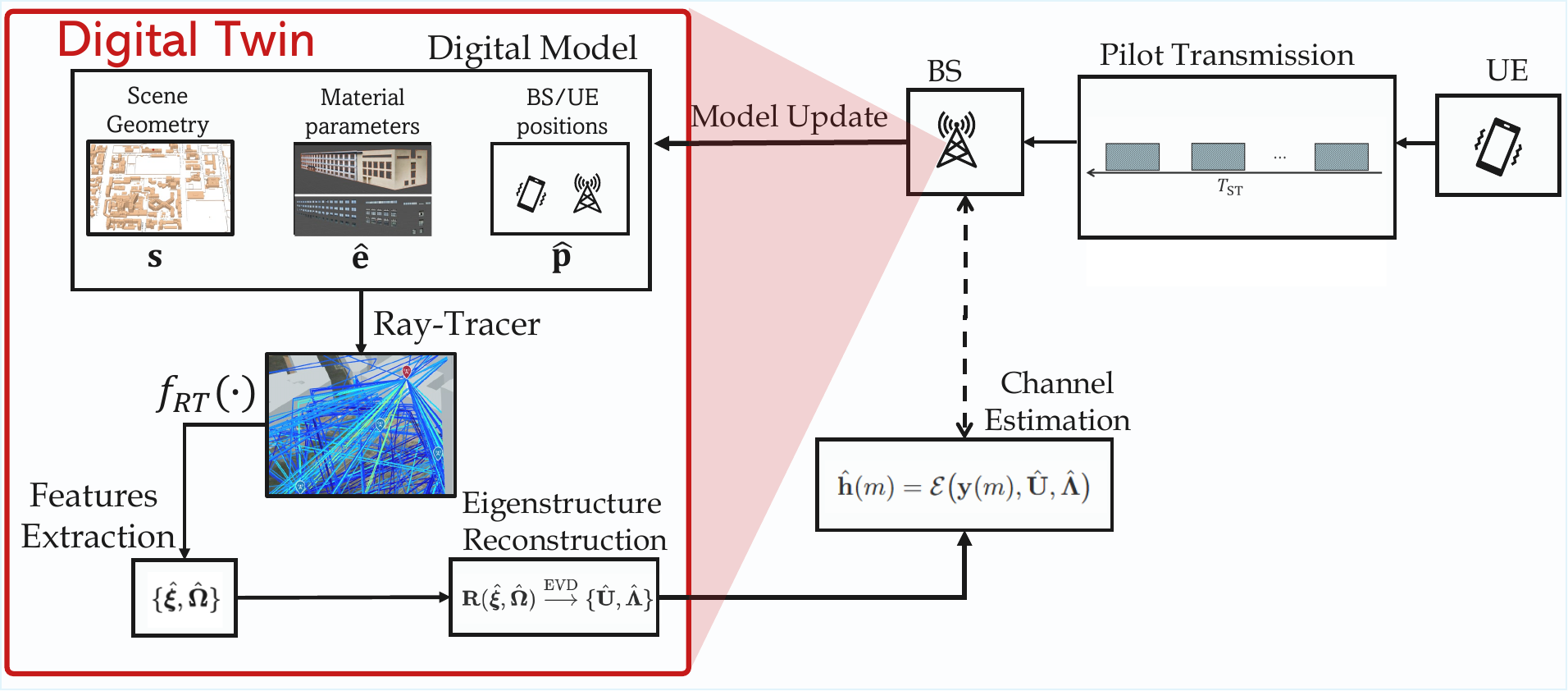}
\caption{DT-empowered channel estimation framework. The DT combines a digital replica—defined by scene geometry $\mathbf{s}$, EM parameters $\hat{\mathbf{e}}$, and BS/UE positions $\hat{\mathbf{p}}$—with RT to estimate propagation parameters $\{\hat{\boldsymbol{\xi}},\hat{\boldsymbol{\Omega}}\}$ and reconstruct the covariance $\mathbf{R}(\hat{\boldsymbol{\xi}},\hat{\boldsymbol{\Omega}})$ and its eigenstructure $\{\hat{\mathbf{U}},\hat{\boldsymbol{\Lambda}}\}$. During uplink training over $T_{\mathrm{ST}}$, the BS combines pilot observations $\mathbf{y}(m)$ with this eigenstructure via $\mathcal{E}(\cdot)$ to obtain the channel estimate.}
\label{Fig: DT framework}
\end{figure}

The DT at the BS estimates the channel parameters set
$\{\hat{\boldsymbol{\xi}},\hat{\boldsymbol{\Omega}}\}$ by combining a digital replica of the propagation environment with RT simulation, as illustrated in Fig.~\ref{Fig: DT framework}. This representation is central to our analysis. It makes explicit how the physical description of the environment is translated into the multipath features that determine the covariance eigenstructure used for channel estimation. The digital replica is defined by the scene geometry $\hat{\mathbf{s}}$, the EM material parameters $\hat{\mathbf{e}}$, and the BS/UE positions $\hat{\mathbf{p}} = [\hat{\mathbf{p}}_\mathrm{BS}^\mathrm{T}, \hat{\mathbf{p}}_\mathrm{UE}^\mathrm{T}]^\mathrm{T}$, representing centroid positions of the arrays in the GCS. Under the planar wave assumption, the AoA, AoD, and delay of each propagation path are common across antenna elements, while the spatial variations across the array are captured by the array response. Accurate digital representations of the environment can be obtained from site-specific maps or 3D reconstruction techniques~\cite{Hoydis_RT}. 

The DT operates in synchronization with the channel evolution over the interval $T_{\mathrm{ST}}$, during which $M$ training symbols and $\{\boldsymbol{\xi},\boldsymbol{\Omega}\}$ are assumed invariant. As illustrated in Fig.~\ref{Fig: DT framework}, at the beginning of each interval $T_{\mathrm{ST}}$, the digital replica, defined by the set $\{\mathbf{s},\hat{\mathbf{e}},\hat{\mathbf{p}}\}$, is used as input to the RT simulation $f_{\mathrm{RT}}(\cdot)$ to generate the estimated channel parameters set $\{\hat{\boldsymbol{\xi}},\hat{\boldsymbol{\Omega}}\}$. The latency and computational requirements of RT-based features generation, scene reconstruction, and dynamic synchronization have been benchmarked in our previous work~\cite{Zhu_RT}, where GPU-accelerated SBR implementations were shown to achieve wall-clock runtimes essentially independent of the interaction depth in multi-link configurations, with the RT cost scaling as $\mathcal{O}(\eta L^{\alpha})$, where $\eta$ is a scenario-dependent parameter and $\alpha\approx 1$ for modern GPU-accelerated implementations. This ensures that, in typical site-specific deployments, the RT runtime satisfies $T_{\mathrm{RT}}\ll T_{\mathrm{ST}}$. Under tighter latency budgets, e.g., high UE mobility, the RT may instead resolve only a reduced subset $\hat{P}<P$ of paths. We addressed this regime in~\cite{DelMoro_DT_2}, whereas here we assume $\hat{P}=P$. Consequently, the propagation parameters can be computed in parallel with the uplink training phase and are assumed constant during the $M$ training symbols.

From the DT-provided parameter sets $\{\hat{\boldsymbol{\xi}},\hat{\boldsymbol{\Omega}}\}$, the channel covariance matrix $\mathbf{R}(\hat{\boldsymbol{\xi}},\hat{\boldsymbol{\Omega}})
=
\mathbf{T}(\hat{\boldsymbol{\xi}})\,\mathrm{diag}(\hat{\boldsymbol{\Omega}})\,\mathbf{T}(\hat{\boldsymbol{\xi}})^\mathsf{H}$ can be reconstructed. The eigenvalue decomposition of $\mathbf{R}(\hat{\boldsymbol{\xi}},\hat{\boldsymbol{\Omega}})$ yields $\mathbf{R}(\hat{\boldsymbol{\xi}},\hat{\boldsymbol{\Omega}})
=
\hat{\mathbf{U}}\hat{\boldsymbol{\Lambda}}\hat{\mathbf{U}}^\mathsf{H}$, where $\hat{\boldsymbol{\Lambda}}=\mathrm{diag}(\hat{\lambda}_1,\dots,\hat{\lambda}_{\hat{r}})$ contains the non-zero eigenvalues and $\hat{\mathbf{U}}=[\hat{\mathbf{u}}_1,\dots,\hat{\mathbf{u}}_{\hat{r}}]$ contains the associated eigenvectors. These eigenvectors define the DT-provided channel subspace $\hat{\mathcal{S}}=\mathrm{span}\{\hat{\mathbf{U}}\}=\mathrm{span}\{\mathbf{T}(\hat{\boldsymbol{\xi}})\}$ capturing the $\hat r$ dominant ST propagation modes. The estimated rank $\hat{r}$ corresponds to the number of dominant eigenvalues of $\mathbf{R}(\hat{\boldsymbol{\xi}},\hat{\boldsymbol{\Omega}})$ and can be determined by standard thresholding rank-selection procedures. 

During the uplink training phase, the UE transmits $M$ pilot symbols while the BS collects the pilot observations $\mathbf{y}(m)$ in~\eqref{eq: Received pilot samples (2)}. The DT provides the structural prior information on the channel subspace and is used to construct projection matrices that exploit the LR structure of the channel covariance. This is the key interface between the DT and channel estimation: mismatches in the digital representation are not used directly but are conveyed through the reconstructed eigenstructure, which precisely determines the estimation performance analyzed later. The DT-empowered channel estimate can therefore be expressed, in a general form, as $\hat{\mathbf h}(m)=\mathcal{E}\big(\mathbf y(m),\hat{\mathbf{U}},\hat{\boldsymbol{\Lambda}}\big)$, where $\mathcal{E}(\cdot)$ denotes an estimator combining the pilot observations with the DT-provided eigenstructure as described below.

\subsection{Digital Twin-based LR and MMSE Channel Estimation}
\label{DT_subsec_estimation}

We first provide the DT-empowered LR estimator. This method has been widely investigated in the context of multi-slot channel estimation~\cite{Mizmizi_V2X, Brighente_RR, Xie_sparsity}, and exploits the invariance of the ST features to project the ML estimate onto the subspace spanned by the dominant propagation modes of the ST MIMO channel. For the DT-empowered LR estimator, the projection matrix, precomputed from the DT-provided covariance, is
\begin{equation}
\label{eq: projection matrix LR DT}
\hat{\mathbf{\Pi}}_\mathrm{LR} = \hat{\mathbf{U}}\hat{\mathbf{U}}^\mathsf{H},
\end{equation}
where $\hat{\mathbf{\Pi}}_\mathrm{LR} \in \mathbb{C}^{L \times L}$ is idempotent and Hermitian. The estimator is then $\hat{\mathbf h}(m)=\mathcal{E}\big(\mathbf y(m),\hat{\mathbf{U}}\big)$, explicitly
\begin{equation}
\label{eq: DT LR estimate}
\hat{\mathbf{h}}_\mathrm{LR}(m) = \hat{\mathbf{\Pi}}_\mathrm{LR}\hat{\mathbf{h}}_\mathrm{ML}(m).
\end{equation}
The LR estimator depends only on the DT-provided channel subspace and is therefore insensitive to perturbations affecting only the eigenvalue profile. This property is central to the paper's scope, as it suggests that geometric mismatches and EM calibration mismatches may not affect LR estimation in the same way. The computation of $\hat{\mathbf{\Pi}}_\mathrm{LR}$ requires the EVD of $\mathbf{R}(\hat{\boldsymbol{\xi}},\hat{\boldsymbol{\Omega}})$, whose cost is $O(L^3)$ in general. Although this operation may become demanding for large values of $L$, it is performed only once per interval $T_{\mathrm{ST}}$ rather than per training symbol. As a result, the DT-empowered LR channel estimation entails first computing $\hat{\mathbf{h}}_\mathrm{ML}(m)$ from the pilot observations and then projecting it onto the DT-provided channel subspace through~\eqref{eq: DT LR estimate}. The overall computational cost is $O\big(LN_{\mathrm{r}}N_\mathrm{p}+L\hat{r}\big)$. The first term corresponds to the computation of the ML channel estimate, whereas the second term accounts for the projection onto the $\hat r$-dimensional DT-provided subspace.

Unlike the DT-empowered LR method, the DT-empowered MMSE estimator requires a weighted projection matrix given by~\cite{Bjornsson_MMSE, Bacci_MMSE}
\begin{equation}
\label{eq: projection matrix MMSE DT}
\hat{\mathbf{\Pi}}_\mathrm{MMSE} =
\hat{\mathbf{U}}
\Big(
\hat{\mathbf{\Lambda}}
\big(
\hat{\mathbf{\Lambda}}
+
\frac{\hat{\sigma}^2_w}{\nu}
\mathbf{I}_{\hat{r}}
\big)^{-1}
\Big)
\hat{\mathbf{U}}^\mathsf{H}.
\end{equation}
The noise power estimate $\hat{\sigma}^2_w$ is obtained as
\begin{equation}
\label{eq: Estimated MMSE noise}
\hat{\sigma}^2_w =
\frac{\nu}{M_w(L-\hat{r})}
\sum_{m=1}^{M_w}
\|
\hat{\mathbf{\Pi}}^\perp_\mathrm{LR}
\hat{\mathbf{h}}_\mathrm{ML}(m)
\|^2,
\end{equation}
where $\hat{\mathbf{\Pi}}^\perp_\mathrm{LR}
=
\mathbf{I}_{L}
-
\hat{\mathbf{\Pi}}_\mathrm{LR}$ is the projection matrix associated with the orthogonal complement of the DT-provided channel subspace. This estimator is accurate when the DT-provided subspace closely matches the true channel subspace. However, subspace mismatches may bias the noise power estimate. Accordingly, the DT-empowered MMSE estimator can be expressed as $\hat{\mathbf h}(m)
=
\mathcal{E}\big(\mathbf y(m),\hat{\mathbf{U}},\hat{\boldsymbol{\Lambda}}\big)$ and is given by
\begin{equation}
\label{eq: DT MMSE}
\hat{\mathbf{h}}_\mathrm{MMSE}(m)
=
\hat{\mathbf{\Pi}}_\mathrm{MMSE}
\hat{\mathbf{h}}_\mathrm{ML}(m).
\end{equation}
The MMSE estimator depends on both the channel subspace and the eigenvalue profile, and is therefore sensitive to both subspace perturbations and eigenvalue distortions. This distinction will be one of the main outcomes of the perturbation and NMSE analysis developed in the next section. In addition, it requires explicit noise power estimation, which adds a preprocessing step before projection. The computational cost remains $O\big(LN_{\mathrm{r}}N_\mathrm{p}+L\hat{r}\big)$, identical to the LR case, since the weights depend only on $\hat{\mathbf{\Lambda}}$ and $\hat{\sigma}_w^2$. In addition, the inversion of the $\hat{r}\times\hat{r}$ matrix in~\eqref{eq: projection matrix MMSE DT} entails a one-time cost $O(\hat{r}^3)$ over the interval $T_{\mathrm{ST}}$. Additionally, $M_w$ OFDM symbols are required for noise power estimation. In practice, $M_w$ can be kept small (typically $M_w \ll M$) without significantly affecting estimation accuracy, thus limiting the associated latency overhead~\cite{Bjornsson_MMSE}.

\subsection{DT Mismatch Modeling}

As detailed in Sec.~\ref{sec:DT_perturbations_metrics_nmse}, we assume that the scene geometry and the BS position are perfectly known, i.e., we consider $\hat{\mathbf{s}}=\mathbf{s}$, and $\hat{\mathbf{p}}_{\mathrm{BS}}=\mathbf{p}_{\mathrm{BS}}$. 
This assumption is consistent with current site-specific reconstruction techniques, which can provide accurate digital representations of the propagation environment~\cite{SionnaRT2023}. In contrast, the UE position and the EM material parameters may be affected by localization and calibration inaccuracies $\hat{\mathbf{e}}=\mathbf{e}+\Delta\mathbf{e}$, and $\hat{\mathbf{p}}_{\mathrm{UE}}=\mathbf{p}_{\mathrm{UE}}+\Delta\mathbf{p}$, where $\Delta\mathbf{e} \sim f_{\Delta\mathbf{e}}(\cdot)$ and
$\Delta\mathbf{p} \sim f_{\Delta\mathbf{p}}(\cdot)$ account for the lack of knowledge of the EM material properties (calibration errors) and UE positioning inaccuracies, respectively (see Sec.\ref{sec:DT_perturbations_metrics_nmse}). In particular, UE positioning errors introduce geometric inconsistencies in the effective site representation seen by the DT and are therefore expected to perturb the propagation paths reconstructed by RT. 

The digital replica is then used for RT simulations, where EM propagation is approximated by launching rays with a prescribed angular resolution and retaining those that satisfy a reception condition~\cite{SionnaRT2023}. Accordingly, the RT simulation extracts estimates of a finite set of $\hat{P}$ dominant multipath components associated with the propagation environment. We denote by $f_{\mathrm{RT}}(\cdot)$ the RT function which outputs the multipath components:
\begin{equation}
f_{\mathrm{RT}}(\mathbf{s},\hat{\mathbf{e}},\hat{\mathbf{p}})
= (\hat{\boldsymbol{\xi}},\hat{\boldsymbol{\Omega}}).
\end{equation}
Remarkably, RT simulations introduce modeling errors due to finite angular resolution or limited reflection order. However, site-specific RT has been shown to accurately reproduce the dominant multipath components in realistic propagation environments.
Therefore, in this work, residual RT approximation errors are neglected in order to isolate the impact of DT mismatches due to localization and EM calibration inaccuracies. This assumption is consistent with sufficiently fine RT simulation settings, for which discretization and finite-resolution effects have been shown to remain limited~\cite{cazzella2025high}. Moreover, the analysis is restricted to the regime of small residual mismatches for which the set of dominant propagation paths remains unchanged under the considered perturbations, so that $\hat{P}=P$. For sufficiently large positioning or calibration errors, the assumption $\hat{P}=P$ may no longer hold, leading to an underestimation or overestimation of the dominant propagation components reconstructed by the DT. This would result in incomplete or inaccurate prior propagation knowledge, with a direct impact on the reconstructed eigenstructure and the resulting channel estimation performance. Such conditions lie outside the scope of the proposed framework and have been investigated in~\cite{DelMoro_DT_2}. This local regime is precisely the one of interest to understand whether DT-empowered channel estimation remains reliable under realistic, yet non-catastrophic, imperfections of the digital representation. 

Accordingly, a first-order approximation around the nominal inputs
$(\mathbf{s},\mathbf{e},\mathbf{p})$ yields
\begin{equation}
f_{\mathrm{RT}}(\mathbf{s},\hat{\mathbf{e}},\hat{\mathbf{p}})
\approx
(\boldsymbol{\xi}+\Delta\boldsymbol{\xi},\boldsymbol{\Omega}+\Delta\boldsymbol{\Omega}),
\end{equation}
where the perturbations $\Delta\boldsymbol{\xi}$ and $\Delta\boldsymbol{\Omega}$ are induced by the perturbations $\Delta\mathbf{p}$ and $\Delta\mathbf{e}$ in the input variables. As mentioned above, we focus on the local perturbation regime. In this context, UE positioning errors primarily affect the propagation geometry and, consequently, the delays and AoA/AoD features, whereas residual EM material calibration errors primarily affect the reflection coefficients and, consequently, the path gains. Therefore, in the considered setting, the cross-sensitivity between geometric and EM parameters is neglected, and we write $\partial \boldsymbol{\xi}/\partial \mathbf{e} = \mathbf{0}$ and $\partial \boldsymbol{\Omega}/\partial \mathbf{p} = \mathbf{0}$\footnote{While the relation $\partial\boldsymbol{\xi}/\partial\mathbf{e}=\mathbf{0}$ holds exactly under specular reflection, since the EM material parameters do not affect the propagation geometry, the assumption $\partial\boldsymbol{\Omega}/\partial\mathbf{p}=\mathbf{0}$ is instead a first-order modeling choice adopted for analytical tractability. In practice, weak position-to-gain coupling may arise through small variations of the propagation distance and local incidence angles. Accounting for such effects would introduce additional gain perturbations following the same analytical pathway as EM calibration errors, without modifying the analytical framework developed in this paper.} This separation is instrumental for the rest of the paper, since it enables us to distinguish how geometric and EM mismatches affect different components of the covariance eigenstructure and, ultimately, different channel estimators. Accordingly, the perturbations can be expressed as
\begin{equation}
\Delta\boldsymbol{\xi}=
\left(\frac{\partial \boldsymbol{\xi}}{\partial \mathbf{p}}\right)\Delta\mathbf{p},
\qquad
\Delta\boldsymbol{\Omega}=
\left(\frac{\partial \boldsymbol{\Omega}}{\partial \mathbf{e}}\right)\Delta\mathbf{e},
\end{equation}
whose statistical characterization will be derived in Sec.~\ref{sec:DT_perturbations_metrics_nmse}.

\section{DT Perturbation Models and NMSE Analysis}

\label{sec:DT_perturbations_metrics_nmse}
\begin{figure}
\centering
\includegraphics[width=0.7\columnwidth]{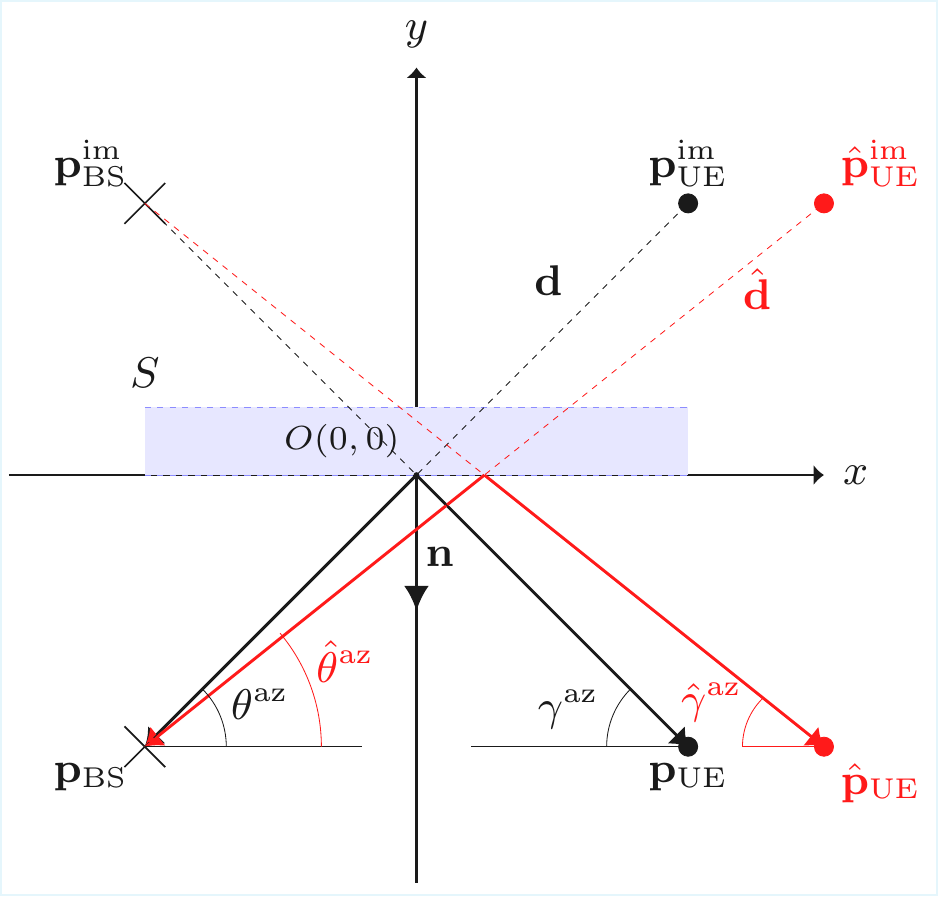}
\caption{Representation of a specular reflection via the image method on a surface $\mathcal{S}$ with normal $\mathbf{n}$. The reflected path is modeled as a LoS link between the BS and the virtual UE image (or equivalently, the virtual BS image and the UE). UE positioning errors propagate through the affine mapping, inducing correlated perturbations in delay and AoA/AoD.}
\label{Fig: Image method}
\end{figure}

In this section, we characterize the perturbations affecting the DT-extracted channel parameter set $\{\hat{\boldsymbol{\xi}},\hat{\boldsymbol{\Omega}}\}$ due to UE positioning and material calibration inaccuracies. Based on this characterization, we then present an NMSE analysis of the DT-empowered estimators, highlighting the role of eigenstructure mismatch. This is the core analytical link of the paper: it shows how DT inaccuracies propagate to the multipath parameters, how they perturb the reconstructed covariance eigenstructure, and why this leads to different robustness properties for LR and MMSE channel estimation. The analysis focuses on the small-error regime, in which the set of dominant propagation paths remains unchanged, and the induced deviations are accurately captured by first-order approximations.

\subsection{Geometry-Induced ST Feature Perturbations}
\label{sec:geometry_uncertainty}

Following the first-order approximation introduced in Sec.~\ref{DT_subsec_generation}, we first characterize the perturbations induced by UE positioning inaccuracies, assuming $\Delta\mathbf{p}\sim\mathcal{N}(\mathbf{0},\boldsymbol{\Sigma}_{\mathbf{p}})$. UE positioning errors introduce geometric mismatches in the effective site representation used by the DT, thereby directly perturbing the propagation paths reconstructed by RT. As illustrated in Fig.~\ref{Fig: Image method}, we adopt the image method~\cite{SionnaRT2023} and focus on specular reflections~\cite{Fuschini2017}. The $p$th propagation path, involving $K_p$ specular reflections, can be equivalently represented as a line-of-sight (LoS) vector between the BS position $\mathbf{p}_{\mathrm{BS}}$ and a virtual UE image position $\mathbf{p}^{\mathrm{im}}_{\mathrm{UE},p}\in\mathbb{R}^3$, or, equivalently, between the UE position $\mathbf{p}_{\mathrm{UE}}$ and the corresponding virtual BS image position $\mathbf{p}^{\mathrm{im}}_{\mathrm{BS},p}\in\mathbb{R}^3$. For the $p$th path, the UE image position is obtained through successive affine reflections and can be written as $\mathbf{p}^{\mathrm{im}}_{\mathrm{UE},p} = \mathbf{A}^{\mathrm{UE}}_p\,\mathbf{p}_{\mathrm{UE}} + \mathbf{b}^{\mathrm{UE}}_p$. Here, $\mathbf{A}^{\mathrm{UE}}_p\in\mathbb{R}^{3\times3}$ is the product of the Householder reflection matrices associated with the reflecting planes along path $p$, and $\mathbf{b}^{\mathrm{UE}}_p\in\mathbb{R}^3$ collects the accumulated affine offsets. Due to the affine structure of the mapping, the perturbed image position satisfies $\hat{\mathbf{p}}^{\mathrm{im}}_{\mathrm{UE},p} = \mathbf{p}^{\mathrm{im}}_{\mathrm{UE},p} + \mathbf{A}^{\mathrm{UE}}_p \Delta\mathbf{p}$. Thus the reflected position error $\Delta\mathbf{p}^{\mathrm{im}}_{p} = \mathbf{A}^{\mathrm{UE}}_p \Delta\mathbf{p}$ is distributed as $\Delta\mathbf{p}^{\mathrm{im}}_{p}\sim\mathcal{N}\!\left(\mathbf{0}, \mathbf{A}^{\mathrm{UE}}_p \boldsymbol{\Sigma}_{\mathbf{p}} \mathbf{A}^{\mathrm{UE}\,\mathsf T}_p \right)$.

We now characterize the impact of UE positioning errors on the delay and AoA features of the $p$th path. Let $\mathbf{r}_p=\mathbf{p}^{\mathrm{im}}_{\mathrm{UE},p}-\mathbf{p}_{\mathrm{BS}}=[r_{p,x}, r_{p,y}, r_{p,z}]^\mathsf{T}$
denote the equivalent LoS vector from the BS to the true UE image position, with norm $R_p=\|\mathbf{r}_p\|$ and unit direction $\mathbf{v}_p^{\mathrm{BS}}=\mathbf{r}_p/R_p$. The latter coincides with the AoA unit vector used in \eqref{array_response_BS}. Under UE positioning uncertainty, the equivalent perturbed vector becomes $\hat{\mathbf{r}}_p=\mathbf{r}_p+\Delta\mathbf{p}^{\mathrm{im}}_p=[\hat r_{p,x},\hat r_{p,y},\hat r_{p,z}]^{\mathsf T}$, and we define $\hat{R}_p=\|\hat{\mathbf{r}}_p\|$. By first-order approximation (see Appendix~\ref{Appendix:B}), the perturbed distance is \begin{equation} \hat{R}_p \approx R_p + (\mathbf{v}_p^{\mathrm{BS}})^{\mathsf T}\Delta\mathbf{p}^{\mathrm{im}}_p. \end{equation} Since $\hat{\tau}_p=\hat{R}_p/c$, it follows that $\hat{\tau}_p\approx\tau_p+\frac{1}{c}(\mathbf{v}_p^{\mathrm{BS}})^\mathsf T\Delta\mathbf{p}^{\mathrm{im}}_p$, so that the propagation delay of the $p$th path is distributed as \begin{equation} \hat{\tau}_p\sim\mathcal{N}\!\left( \tau_p, \frac{1}{c^2} (\mathbf{v}_p^{\mathrm{BS}})^{\mathsf T} \mathbf{A}^{\mathrm{UE}}_p \boldsymbol{\Sigma}_{\mathbf{p}} \mathbf{A}^{\mathrm{UE}\,\mathsf T}_p \mathbf{v}_p^{\mathrm{BS}} \right). \end{equation} 
The azimuth and elevation AoA features follow from the same first-order principle (see Appendix~\ref{Appendix:B}). Defining $\rho_p^2=r_{p,x}^2+r_{p,y}^2$, the azimuth AoA $\hat{\theta}_p^{\mathrm{az}}=\mathrm{atan2}(\hat{r}_{p,y},\hat{r}_{p,x})$ can be approximated as
\begin{equation} \hat{\theta}_p^{\mathrm{az}} \approx \theta_p^{\mathrm{az}} + \frac{1}{\rho_p^2} \left(\mathbf{q}^{\theta_{\mathrm{az}}}_p\right)^{\mathsf T} \Delta\mathbf{p}^{\mathrm{im}}_p, \end{equation}
where $\mathbf{q}^{\theta_{\mathrm{az}}}_p=\left[-r_{p,y},\,r_{p,x},\,0\right]^{\mathsf T}$. Accordingly, the azimuth AoA of the $p$th path is distributed as \begin{equation} \hat{\theta}_p^{\mathrm{az}} \sim \mathcal{N}\!\left( \theta_p^{\mathrm{az}}, \frac{1}{\rho_p^4} \left(\mathbf{q}^{\theta_{\mathrm{az}}}_p\right)^\mathsf T \mathbf{A}^{\mathrm{UE}}_p \boldsymbol{\Sigma}_{\mathbf{p}} \mathbf{A}^{\mathrm{UE}\,\mathsf T}_p \mathbf{q}^{\theta_{\mathrm{az}}}_p \right).
\end{equation}
Similarly, the elevation AoA $\hat{\theta}_p^{\mathrm{el}}=\arcsin(\hat{r}_{p,z}/\hat{R}_p)$ can be approximated as 
\begin{equation} \hat{\theta}_p^{\mathrm{el}} \approx \theta_p^{\mathrm{el}} + \frac{1}{R_p^2\rho_p} \left(\mathbf{q}^{\theta_{\mathrm{el}}}_p\right)^{\mathsf T} \Delta\mathbf{p}^{\mathrm{im}}_p, \end{equation}
where $\mathbf{q}^{\theta_{\mathrm{el}}}_p=\left[-r_{p,z}r_{p,x},\,-r_{p,z}r_{p,y},\,\rho_p^2\right]^{\mathsf T}$. This yields an elevation AoA, for the $p$th path, distributed as \begin{equation} \hat{\theta}_p^{\mathrm{el}} \sim \mathcal{N}\!\left( \theta_p^{\mathrm{el}}, \frac{1}{R_p^4\rho_p^2} \left(\mathbf{q}^{\theta_{\mathrm{el}}}_p\right)^\mathsf T \mathbf{A}^{\mathrm{UE}}_p \boldsymbol{\Sigma}_{\mathbf{p}} \mathbf{A}^{\mathrm{UE}\,\mathsf T}_p \mathbf{q}^{\theta_{\mathrm{el}}}_p \right).
\end{equation}

An analogous construction applies to the AoD features. Using the image method, the same propagation path can be equivalently represented as a LoS vector from the UE to the BS image position. The BS image position is built in the same way as the UE image position, but following the same sequence of reflections in reverse order. Let $\mathbf{d}_p=\mathbf{p}_{\mathrm{UE}}-\mathbf{p}^{\mathrm{im}}_{\mathrm{BS},p}=[d_{p,x}, d_{p,y}, d_{p,z}]^{\mathsf T}$ denote the corresponding vector, with norm $D_p=\|\mathbf{d}_p\|$ and unit direction $\mathbf{v}_p^{\mathrm{UE}}=\mathbf{d}_p/D_p$, which coincides with the AoD unit vector used in \eqref{array_response_UE}. Under UE positioning uncertainty, the perturbed vector is $\hat{\mathbf{d}}_p=\mathbf{d}_p+\Delta\mathbf{p}$. Defining $\delta_p^2=d_{p,x}^2+d_{p,y}^2$ and applying a first-order approximation of the angular features, the azimuth and elevation AoDs are Gaussian distributed as
\begin{equation}
\hat{\gamma}_p^{\mathrm{az}} \sim \mathcal{N}\!\left(
\gamma_p^{\mathrm{az}}, 
\frac{1}{\delta_p^4}
\left(\mathbf{q}^{\gamma_{\mathrm{az}}}_p\right)^\mathsf T 
\boldsymbol{\Sigma}_{\mathbf{p}} 
\mathbf{q}^{\gamma_{\mathrm{az}}}_p
\right),
\end{equation}
\begin{equation}
\hat{\gamma}_p^{\mathrm{el}} \sim \mathcal{N}\!\left(
\gamma_p^{\mathrm{el}}, 
\frac{1}{D_p^4\delta_p^2}
\left(\mathbf{q}^{\gamma_{\mathrm{el}}}_p\right)^\mathsf T 
\boldsymbol{\Sigma}_{\mathbf{p}} 
\mathbf{q}^{\gamma_{\mathrm{el}}}_p
\right),
\end{equation}
where $\mathbf{q}^{\gamma_{\mathrm{az}}}_p=\left[-d_{p,y},\,d_{p,x},\,0\right]^{\mathsf T}$ and $\mathbf{q}^{\gamma_{\mathrm{el}}}_p=\left[-d_{p,z}d_{p,x},\,-d_{p,z}d_{p,y},\,\delta_p^2\right]^{\mathsf T}$.
Under the adopted first-order approximations, the DT-provided ST features of the $p$th path can be written as $\hat{\boldsymbol{\xi}}_p = \boldsymbol{\xi}_p + \Delta\boldsymbol{\xi}_p$, where $\Delta\boldsymbol{\xi}_p = \mathbf{J}_p\Delta\mathbf{p}$ is distributed as $\Delta\boldsymbol{\xi}_p\sim\mathcal{N}\!\left(\mathbf{0}, \mathbf{J}_p \boldsymbol{\Sigma}_{\mathbf{p}} \mathbf{J}_p^{\mathsf T} \right)$, with $\mathbf{J}_p\in\mathbb{R}^{5\times 3}$ given by
\begin{equation}
\mathbf{J}_p =
\begin{bmatrix}
\frac{1}{\rho_p^2}\left(\mathbf{q}^{\theta_{\mathrm{az}}}_p\right)^{\mathsf T}\mathbf{A}^{\mathrm{UE}}_p \\[6pt]
\frac{1}{R_p^2\rho_p}\left(\mathbf{q}^{\theta_{\mathrm{el}}}_p\right)^{\mathsf T}\mathbf{A}^{\mathrm{UE}}_p \\[6pt]
\frac{1}{\delta_p^2}\left(\mathbf{q}^{\gamma_{\mathrm{az}}}_p\right)^{\mathsf T} \\[6pt]
\frac{1}{D_p^2\delta_p}\left(\mathbf{q}^{\gamma_{\mathrm{el}}}_p\right)^{\mathsf T} \\[6pt]
\frac{1}{c}(\mathbf{v}_p^{\mathrm{BS}})^{\mathsf T}\mathbf{A}^{\mathrm{UE}}_p
\end{bmatrix}.
\end{equation}
Collecting the $P$ multipath components, the DT-provided ST features are $\hat{\boldsymbol{\xi}}=\boldsymbol{\xi}+\Delta\boldsymbol{\xi}$, with $\Delta\boldsymbol{\xi}\sim\mathcal{N}\!\left(\mathbf{0},\mathbf{J}\boldsymbol{\Sigma}_{\mathbf{p}}\mathbf{J}^{\mathsf T}\right)$ and
$\mathbf{J}=[\mathbf{J}_1^{\mathsf T},\dots,\mathbf{J}_P^{\mathsf T}]^{\mathsf T}\in\mathbb{R}^{5P\times3}$. This result makes explicit how UE positioning uncertainty, through its geometric effect on the site representation, perturbs the ST features that define the channel subspace.

\subsection{Calibration-Induced Path Gains Perturbations}
\label{sec:material_uncertainty}

Following the same first-order approximation introduced in Sec.~\ref{DT_subsec_generation}, we now model the perturbations induced by uncertainties in the EM material parameters. Recall that, under this approximation, UE positioning errors perturb only the propagation geometry, i.e., the delays and AoA/AoD features, while their impact on the path gains is neglected. This complementary perturbation model isolates the effect of EM mismatches, whose main impact is expected on the eigenvalue profile rather than on the channel subspace. Consider the $p$th propagation path involving $K_p$ successive specular reflections. The path gain extracted from RT can be written as
\begin{equation} 
\label{eq: path gain mult} \hat{\Omega}_p = \Omega^{(0)}_p \prod_{k=1}^{K_p} \big| \hat{\Gamma}_{p,k} \big|^2, 
\end{equation} 
where $\Omega^{(0)}_p>0$ collects the free-space attenuation contribution, and $\hat{\Gamma}_{p,k}\in\mathbb{C}$ denotes the reflection coefficient used by the digital model for the $k$th interaction along path $p$. For the $k$th reflection, let $\mathbf{k}_{p,k}\in\mathbb{R}^3$ denote the unit incident propagation direction and $\mathbf{n}_{p,k}\in\mathbb{R}^3$ the unit outward normal of the corresponding planar surface. The local incidence angle $\theta_{p,k}\in[0,\pi/2]$ is defined with respect to the surface normal as $\cos(\theta_{p,k}) = -\mathbf{k}_{p,k}^{\mathsf T}\mathbf{n}_{p,k}$. The reflection coefficient depends on the complex relative permittivity $\varepsilon^{r}_{p,k}\in\mathbb{C}$ of the $k$th reflecting surface, which can be written as $\varepsilon^{r}_{p,k}=\varepsilon'_{p,k}-j\varepsilon''_{p,k}$. The real part $\varepsilon'_{p,k}$ models the dielectric permittivity of the material, while the imaginary part $\varepsilon''_{p,k}$ accounts for conductive and loss effects. The calibrated permittivity used by the DT is modeled as $\hat{\varepsilon}^{r}_{p,k} = \varepsilon^{r}_{p,k} + \Delta e_{p,k}$, where $\varepsilon^{r}_{p,k}\in\mathbb{C}$ is the true relative permittivity and $\Delta e_{p,k}$ is the corresponding calibration error. We denote by $\Delta\mathbf{e}_p=[\Delta e_{p,1},\dots,\Delta e_{p,K_p}]^{\mathsf T}$ the vector collecting the calibration errors along path $p$, and by $\Delta\mathbf{e}$ the vector collecting all material calibration errors in the DT model. The calibration error is modeled as $\Delta e_{p,k}\sim\mathcal{N}_{\mathbb{C}}(0,\sigma_{p,k}^{2})$, with independent perturbations across reflections and paths.

For an air–material interface (incident medium relative permittivity equal to one), the Fresnel reflection coefficients are given by
\begin{align} &\hat{\Gamma}_{p,k}^{\mathrm{TE}} = \frac{\cos(\theta_{p,k}) - \sqrt{\hat{\varepsilon}^{r}_{p,k}-\sin^{2}(\theta_{p,k})}} {\cos(\theta_{p,k}) + \sqrt{\hat{\varepsilon}^{r}_{p,k}-\sin^{2}(\theta_{p,k})}},\\ &\hat{\Gamma}_{p,k}^{\mathrm{TM}} = \frac{\hat{\varepsilon}^{r}_{p,k}\cos(\theta_{p,k}) - \sqrt{\hat{\varepsilon}^{r}_{p,k}-\sin^{2}(\theta_{p,k})}} {\hat{\varepsilon}^{r}_{p,k}\cos(\theta_{p,k}) + \sqrt{\hat{\varepsilon}^{r}_{p,k}-\sin^{2}(\theta_{p,k})}}.
\end{align}
In the following, $\Gamma_{p,k}$ denotes either the TE or TM coefficient, or an equivalent scalar coefficient capturing the effective polarization state along the considered path. Writing explicitly $\Gamma_{p,k}=\Gamma(\varepsilon^r_{p,k},\theta_{p,k})$ and assuming small calibration errors, a first-order approximation yields
\begin{equation} 
\hat{\Gamma}_{p,k} \approx \Gamma_{p,k} +\underbrace{\left. \frac{\partial \Gamma(\varepsilon,\theta_{p,k})}{\partial \varepsilon} \right|_{\varepsilon=\varepsilon^r_{p,k}}}_{c_{p,k}}\,\Delta e_{p,k}, 
\end{equation} 
where the sensitivity factor $c_{p,k}$ is derived in Appendix~\ref{Appendix:C}. Taking the squared magnitude gives $ |\hat{\Gamma}_{p,k}|^2 \approx |\Gamma_{p,k}|^2 + 2\,\Re\!\left\{\Gamma_{p,k}^* c_{p,k}\,\Delta e_{p,k}\right\}$, where second-order terms in $\Delta e_{p,k}$ are neglected. Substituting this into \eqref{eq: path gain mult} and approximating the product to first order around the nominal values yields
\begin{equation} \hat{\Omega}_p \approx \Omega_p + \Omega_p\sum_{k=1}^{K_p} \Re\!\left\{ \alpha_{p,k}\, \Delta e_{p,k} \right\}, \end{equation}
where the nominal path gain is $\Omega_p = \Omega^{(0)}_p \prod_{k=1}^{K_p} |\Gamma_{p,k}|^2$ and the sensitivity factor (see Appendix~\ref{Appendix:C}) is $\alpha_{p,k} = 2c_{p,k}/\Gamma_{p,k}$. The perturbed path gain can therefore be written as $\hat{\Omega}_p = \Omega_p + \Delta\Omega_p$, with 
$\Delta\Omega_p = \Omega_p\sum_{k=1}^{K_p} \Re\!\left\{ \alpha_{p,k}\, \Delta e_{p,k} \right\}$.

Since $\alpha_{p,k}\Delta e_{p,k}$ is circularly symmetric complex Gaussian with variance $|\alpha_{p,k}|^2\sigma_{p,k}^2$, its real part is Gaussian with variance $|\alpha_{p,k}|^2\sigma_{p,k}^2/2$. Therefore,
\begin{equation}
\Delta\Omega_p \sim \mathcal{N}\!\left(0,\, \Omega_p^2 \sum_{k=1}^{K_p} \frac{\sigma_{p,k}^2}{2}\, |\alpha_{p,k}|^2 \right).
\end{equation}
Denoting the perturbation vector $\Delta\boldsymbol{\Omega}=[\Delta\Omega_1,\dots,\Delta\Omega_{\hat{P}}]^{\mathsf T}$, it follows that $\Delta\boldsymbol{\Omega}\sim \mathcal{N}\!\left(\mathbf{0},\boldsymbol{\Sigma}_{\boldsymbol{\Omega}}\right)$, where $\boldsymbol{\Sigma}_{\boldsymbol{\Omega}}=\mathrm{diag}(\sigma_{\Omega_1}^2,\dots,\sigma_{\Omega_P}^2)$ with $\sigma_{\Omega_p}^2=\Omega_p^2 \sum_{k=1}^{K_p} \frac{\sigma_{p,k}^2}{2}\, |\alpha_{p,k}|^2$. In the considered regime, EM calibration errors thus redistribute power across the propagation modes without directly modifying the underlying geometric structure.

\subsection{NMSE Analysis}
\label{NMSE Analysis}
We carry out the NMSE analysis by taking the expectation over the channel and noise realizations, assumed independent, while conditioning on the DT-provided eigenstructure to isolate the impact of the specific DT mismatch. In addition, the perturbations associated with UE positioning and EM calibration errors are assumed statistically independent of the receiver noise.\footnote{This assumption follows from the considered DT framework, where UE positioning, calibration, and channel estimation are performed as separate estimation processes using different observations or time instants, thus ensuring statistical independence.} This yields closed-form expressions for the NMSE conditioned on the DT-provided eigenstructure. The aim of the analysis is to make explicit how different perturbations of this eigenstructure translate into distinct degradation mechanisms for the DT-empowered LR and MMSE estimators. Let $\hat{\mathbf h}(m)$ denote the channel estimate and define the estimation error as $\widetilde{\mathbf h}(m)=\hat{\mathbf h}(m)-\mathbf h(m)$. The corresponding NMSE is
\begin{equation}
\label{eq:NMSE_def} \mathrm{NMSE} = \frac{1}{\Gamma} \mathbb{E} \Big\{ \widetilde{\mathbf h}^{\mathsf H}(m) \widetilde{\mathbf h}(m) \Big\}, 
\end{equation}
where $\Gamma=\mathbb{E}\{\|\mathbf{h}(m)\|^2\}=\mathrm{Tr}\{\mathbf{R}(\boldsymbol{\xi},\boldsymbol{\Omega})\}$ is the average channel power.

For the LR case, plugging~\eqref{eq: DT LR estimate} into~\eqref{eq:NMSE_def}, the NMSE admits the decomposition
\begin{equation}
\begin{split} 
\label{eq: MSE LR DT insights} \mathrm{NMSE}^{\mathrm{DT}}_{\mathrm{LR}} 
&= \frac{1}{\Gamma} \Big( \mathrm{Tr}\{\hat{\mathbf{\Pi}}^\perp_\mathrm{LR}\mathbf{R}(\boldsymbol{\xi},\boldsymbol{\Omega})\} + \mathrm{Tr}\{\hat{\mathbf{\Pi}}_\mathrm{LR}\} \frac{\sigma^2_w}{\nu} \Big) \\ 
&= \sigma^2_{h,\mathrm{LR}} + \sigma^2_{w,\mathrm{LR}},
\end{split} 
\end{equation}
where $\hat{\mathbf{\Pi}}^\perp_\mathrm{LR}=\mathbf{I}_L-\hat{\mathbf{\Pi}}_\mathrm{LR}$ projects onto $\hat{\mathcal{S}}^\perp$. The two terms $\sigma^2_{h,\mathrm{LR}}$ and $\sigma^2_{w,\mathrm{LR}}$ represent, respectively, the residual channel component lying outside the DT-provided subspace and the projected noise component. From the properties of the projection matrices, $\mathrm{Tr}\{\hat{\mathbf{\Pi}}_{\mathrm{LR}}\}=\mathrm{rank}\{\hat{\mathbf{\Pi}}_{\mathrm{LR}}\}=\hat{r}$, and thus $\sigma^2_{w,\mathrm{LR}}=\sigma^2_w \hat{r}/(\nu\Gamma)$. For the channel term, using $\mathbf{R}(\boldsymbol{\xi},\boldsymbol{\Omega}) =\sum_{j=1}^{r}\lambda_j\mathbf u_j\mathbf u_j^{\mathsf H}$ and $\hat{\mathbf{\Pi}}_{\mathrm{LR}} =\sum_{k=1}^{\hat r}\hat{\mathbf u}_k\hat{\mathbf u}_k^{\mathsf H}$ yields
\begin{equation}
\label{eq:Channel_projection_LR} 
\sigma^2_{h,\mathrm{LR}} =1-\frac{1}{\Gamma} \sum_{j=1}^{r}\sum_{k=1}^{\hat r} \lambda_j\,|\langle\hat{\mathbf u}_k,\mathbf u_j\rangle|^2 , 
\end{equation}
which directly quantifies the channel energy lying outside the DT-provided subspace. 

Analogously, for the MMSE, substituting~\eqref{eq: DT MMSE} into~\eqref{eq:NMSE_def} gives
\begin{align}
\label{eq: MSE MMSE DT insights} \mathrm{NMSE}^{\mathrm{DT}}_{\mathrm{MMSE}} &\!=\!\frac{1}{\Gamma}\!\mathrm{Tr}\Big\{\!(\mathbf I_L\!-\!\hat{\mathbf\Pi}_\mathrm{MMSE}) \mathbf R(\boldsymbol{\xi},\boldsymbol{\Omega}) (\mathbf I_L\!-\!\hat{\mathbf\Pi}_\mathrm{MMSE})^{\mathsf H}\!\Big\}\nonumber\\
&\quad+\frac{1}{\Gamma}\mathrm{Tr}\!\Big\{\hat{\mathbf\Pi}_\mathrm{MMSE} \hat{\mathbf\Pi}_\mathrm{MMSE}^{\mathsf H}\Big\}\frac{\sigma_w^2}{\nu}\nonumber\\
&\!=\!\sigma^2_{h,\mathrm{MMSE}}+\sigma^2_{w,\mathrm{MMSE}}, 
\end{align}
which again decomposes into a residual channel (distortion) component and a noise component. Unlike the LR case, however, the matrix $\hat{\mathbf{\Pi}}_\mathrm{MMSE}$ is Hermitian but not idempotent. Inserting~\eqref{eq: projection matrix MMSE DT} into~\eqref{eq: MSE MMSE DT insights} gives the noise contribution $ \sigma^2_{w,\mathrm{MMSE}} = (\sigma_w^2/\nu \Gamma) \sum_{k=1}^{\widehat{r}}\mu_k^2$, where $\mu_k=\nu\hat{\lambda}_k/(\nu\hat{\lambda}_k+\hat{\sigma}_w^2)$ is the noise power retained after eigenvalue-dependent shrinkage. Similarly, using the spectral decomposition $\hat{\mathbf{\Pi}}_{\mathrm{MMSE}} =\sum_{k=1}^{\widehat r}\mu_k\hat{\mathbf u}_k\hat{\mathbf u}_k^{\mathsf H}$, the channel term becomes
\begin{equation}
\label{eq:Channel_projection_MMSE} 
\sigma^2_{h,\mathrm{MMSE}} =
1-\frac{1}{\Gamma}
\sum_{j=1}^{r}\sum_{k=1}^{\widehat r}
\lambda_j(2\mu_k-\mu_k^2)
|\langle\hat{\mathbf u}_k,\mathbf u_j\rangle|^2 . 
\end{equation}
As $\sigma_w^2\rightarrow 0$, the shrinkage coefficients satisfy $\mu_k\rightarrow 1$ for $k=1,\dots,\hat{r}$, so that $\hat{\mathbf{\Pi}}_{\mathrm{MMSE}}\rightarrow\hat{\mathbf{\Pi}}_{\mathrm{LR}}$ and the MMSE estimator asymptotically coincides with the LR estimator. These expressions reveal a key difference between the two estimators under eigenstructure mismatch: LR depends only on subspace mismatch, whereas MMSE is also sensitive to eigenvalue distortions through the shrinkage coefficients. 

Note that, when $\hat{r}<r$, fewer dominant modes are retained
and the LR channel term $\sigma^2_{h,\mathrm{LR}}$ increases because a larger portion of the channel energy lies outside the estimated subspace. Conversely, when $\hat{r}>r$, the LR noise term $\sigma^2_{w,\mathrm{LR}}$ increases because
additional noisy dimensions are included in the projection. The MMSE
estimator is affected in both cases through the shrinkage coefficients $\mu_k$, which
nonetheless inherently attenuate noise, providing intrinsic robustness to moderate
rank overestimation.

Let us now assume correct rank-order selection, i.e., $\hat r=r$. We exploit the previous NMSE analysis to assess the impact of EM calibration and positioning perturbations separately. Consider first calibration perturbations only, i.e., $\hat{\boldsymbol{\xi}}=\boldsymbol{\xi}$, for which the covariance matrix is $\hat{\mathbf R}_\Omega=\mathbf T(\boldsymbol{\xi})\mathrm{diag}(\hat{\boldsymbol{\Omega}})\mathbf T^{\mathsf H}(\boldsymbol{\xi})$. Since these perturbations affect only the path gains, the column space of $\mathbf T(\boldsymbol{\xi})$ is preserved and the channel subspace is unchanged, i.e., $\mathrm{span}\{\hat{\mathbf R}_\Omega\}=\mathrm{span}\{\mathbf R(\boldsymbol{\xi},\boldsymbol{\Omega})\}$. The estimated and true subspace dimensions then coincide, so that the LR projection equals the orthogonal projector onto the true channel subspace, giving $\sigma^2_{h,\mathrm{LR}}=0$ and $\sigma^2_{w,\mathrm{LR}}=\sigma^2_w r/(\nu\Gamma)$, consistent with the asymptotic LR behavior in~\cite{Brighente_RR}. The MMSE estimator, by contrast, remains affected: the eigenvalues of the covariance matrix change, which modifies the shrinkage coefficients $\mu_k$ even though the subspace itself is unchanged. 

Consider now positioning perturbations only, i.e., $\hat{\boldsymbol{\Omega}}=\boldsymbol{\Omega}$, for which the ST responses become $\mathbf T(\hat{\boldsymbol{\xi}})$ and the covariance matrix is $\hat{\mathbf R}_{\xi}=\mathbf T(\hat{\boldsymbol{\xi}})\mathrm{diag}(\boldsymbol{\Omega})\mathbf T^{\mathsf H}(\hat{\boldsymbol{\xi}})$. Here the perturbation alters the column space of $\mathbf T(\boldsymbol{\xi})$, so that in general $\mathrm{span}\{\mathbf T(\hat{\boldsymbol{\xi}})\}\neq\mathrm{span}\{\mathbf T(\boldsymbol{\xi})\}$. The LR projection is thus built from a mismatched subspace, yielding $\sigma^2_{h,\mathrm{LR}}>0$, and the MMSE estimator is likewise affected, since both eigenvectors and eigenvalues are perturbed. This formalizes the main insight anticipated above: geometric mismatches due to UE positioning errors are the dominant source of degradation, because they alter the channel subspace, whereas EM calibration errors mainly perturb the eigenvalue profile and therefore have a more limited impact, especially for LR estimation. Notably, this conclusion remains valid even in the presence of weak position-to-gain coupling, which has been previously omitted. Indeed, the resulting gain perturbations affect only the covariance eigenvalue profile. Consequently, the LR estimator remains unaffected, whereas the MMSE estimator may experience only a limited additional degradation through the shrinkage coefficients.

To quantify the magnitude of the impact of positioning and EM calibration errors due to eigenstructure mismatch, we refer to the numerical results presented in the next section.

\section{Numerical Results}
\label{Numerical results}

This section evaluates the DT-empowered estimators under site-specific RT simulations and residual positioning and calibration perturbations, describing the simulation setup and assessing performance against baseline methods.

\subsection{Simulation Framework}
\label{sec: Simulation Framework}

\begin{table}[t!]
\centering
\caption{RT simulation parameters.}
\renewcommand{\arraystretch}{1.1}
\setlength{\tabcolsep}{4pt}
\begin{tabular}{l c l c}
\toprule
\footnotesize
\textbf{Parameter} & \textbf{Value} & \textbf{Parameter} & \textbf{Value} \\
\midrule
Carrier frequency & 28 GHz & Grid UE height & 1.5 m \\
Antenna UE height & 0.2 m & Grid resolution & $2 \times 2$ m$^2$ \\
SUMO sampling time & 0.1 s & RT method & Fibonacci (SBR) \\
Initial sampled rays & $10^6$ & Interaction types & Reflections \\
\bottomrule
\end{tabular}
\label{tab:simulation_parameters}
\end{table}

In this study, we consider three outdoor scenarios with different building densities and propagation characteristics. The reference scenario is the 3D model of an urban area in Milan, Italy, previously investigated in~\cite{cazzella2025high,Zhu_RT}, covering an area of approximately $550 \times 670$ m$^2$, with a single base station positioned on a rooftop at a height of 21.7 m. To achieve accurate RT channel generation, we selected NVIDIA Sionna RT (v0.18.0)~\cite{SionnaRT2023}. We integrated with NVIDIA Sionna the 3D model containing buildings, walls, ground, and parked vehicle meshes. This initial high-fidelity base scenario is enriched with EM material properties assigned to each object through pre-processing and by referring to the ITU recommendation~\cite{ITUR2040}. We refer to this as the \emph{urban} scenario.

To assess the generality of the proposed framework, we additionally consider the \emph{suburban} and \emph{rural} scenarios, generated from open geospatial data provided by the Estonian Land Board portal\footnote{\url{https://geoportaal.maaamet.ee/}.}. The two scenarios correspond to distinct areas of Tallinn, Estonia, each covering $300 \times 300$ m$^2$, with BS heights of $20.5\:\text{m}$ and $21\:\text{m}$ for the suburban and rural cases, respectively. The \emph{suburban} scenario exhibits an intermediate building density, whereas the \emph{rural} scenario is characterized by sparse buildings and reduced multipath richness. The RT settings reported in Table~\ref{tab:simulation_parameters} are kept identical across all three scenarios, so that only the propagation environment changes. Unless otherwise specified, the \emph{urban} scenario is used as the reference case throughout the paper, while the \emph{suburban} and \emph{rural} scenarios are introduced to verify that the main conclusions are not specific to dense urban propagation conditions.

For each environment, two complementary RT datasets are generated: (i) a \textit{grid-based} simulation, used only to build the fingerprinting database, where UEs are placed on a regular spatial grid ($2\times 2$ m$^2$) at fixed height and no vehicular dynamics are included; and (ii) a \textit{vehicular} simulation, used in all performance evaluations as ground-truth channel realization, where UEs are placed on realistic vehicle meshes generated by SUMO~\cite{SUMO2018} and vehicle dynamics are explicitly modeled. In both setups, RT provides, for each UE--BS pair and propagation path, the azimuth and elevation AoA/AoD, the propagation delay, and the path gain.

The adopted MIMO-OFDM system assumes a uniform linear array (ULA) at the UE and a uniform planar array (UPA) at the BS, both with antenna spacing equal to $\lambda/2$. A root-raised cosine (RRC) pulse shaping filter with roll-off factor $\beta = 0.2$ is employed. The system follows the 5G NR numerology $\mu = 4$, corresponding to a subcarrier spacing of 240 kHz. Each slot contains 14 OFDM symbols, with a slot duration of $T_{\mathrm{slot}} = 62.5,\mu$s. In accordance with the adopted 5G NR slot structure, one uplink training OFDM symbol is transmitted per slot. In the following, we consider $M=100$ training symbols, corresponding to an interval $T_{\mathrm{ST}}=6.25$ ms~\cite{Brighente_RR}. Furthermore, we assume a fixed number of transmit antennas $N_\mathrm{t}=2$ and a number of pilot subcarriers $N_\mathrm{p}=N_{\mathrm{sc}}/2$.

For numerical comparison, the following baseline estimators are considered: a ML estimator~\cite{Gao_ML}; Conventional data-driven LR and MMSE estimators~\cite{Brighente_RR,Bacci_MMSE}, where the channel covariance matrix is estimated from $M=100$ ML channel estimates. The rank is set equal to the true channel subspace dimension to isolate the effect of covariance mismatch; Fingerprinting-based LR and MMSE estimators~\cite{Mizmizi_V2X,QIU_CKM}, where the prior ST eigenstructure is retrieved from the grid point closest to the estimated UE position obtained from the \textit{grid-based} simulation; Ideal LR and MMSE estimators, assuming perfect knowledge of the true channel eigenstructure.
For all estimators except ML and ideal ones, the noise power is estimated from $M_w=1$ training symbol as in~\eqref{eq: Estimated MMSE noise}. 

To model EM calibration mismatches in a compact form, we introduce a relative gain perturbation model. This model provides a simplified aggregate representation of EM calibration uncertainties, consistent with the first-order analysis. The considered perturbation levels are not intended to reproduce a specific material estimation error. Rather, they represent moderate aggregate levels of uncertainty affecting the predicted path gains, encompassing inaccuracies in the EM characterization of materials, simplified material descriptions, and calibration imperfections. Specifically, the perturbation affecting the $p$th path is expressed as $\Delta\Omega_p = \Omega_p \beta_p$, where $\beta_p \sim \mathcal{N}(0,\sigma_g^2)$ captures the aggregate effect of EM calibration errors along the propagation path, and $\sigma_g$ denotes the normalized standard deviation. Accordingly, the gain perturbation vector satisfies $\Delta\boldsymbol{\Omega} \sim \mathcal{N}(\mathbf{0},\boldsymbol{\Sigma}_{\boldsymbol{\Omega}})$, with $\boldsymbol{\Sigma}_{\boldsymbol{\Omega}}=\sigma_g^2\,\mathrm{diag}(\boldsymbol{\Omega}^2)$. Positioning uncertainty is modeled as $\Delta\mathbf{p}\sim\mathcal{N}(\mathbf{0},\sigma_p^2\mathbf{I}_3)$, where $\sigma_p$ denotes the standard deviation of the UE positioning error. The resulting ST feature perturbations satisfy $\Delta\boldsymbol{\xi}\sim \mathcal{N}(\mathbf{0},\sigma_p^2\mathbf{J}\mathbf{J}^{\mathsf{T}})$, where the matrix $\mathbf{J}$ (see Sec.~\ref{sec:geometry_uncertainty}) is computed from the scene geometry.

\subsection{Subspace Similarity Analysis}
\label{sec:subspace_similarity_results}

We first assess the impact of positioning and EM calibration errors on the DT-provided channel covariance matrix. To this end, we adopt the similarity metric
\begin{equation}
\eta = \mathbb{E}\left\{ \frac{\mathrm{Tr}\left\{ \mathbf{R}(\boldsymbol{\xi},\boldsymbol{\Omega})^\mathsf{H} \mathbf{R}(\hat{\boldsymbol{\xi}},\hat{\boldsymbol{\Omega}})\right\}}
{\mathrm{Tr}\left\{\mathbf{R}(\boldsymbol{\xi},\boldsymbol{\Omega})^\mathsf{H}\right\} \, \mathrm{Tr}\left\{\mathbf{R}(\hat{\boldsymbol{\xi}},\hat{\boldsymbol{\Omega}})\right\} }\right\},
\end{equation}
which provides a normalized measure of the average correlation between the ground-truth and DT-provided covariance matrices. This metric jointly captures subspace deviations and eigenvalue distortions, and therefore reflects the overall eigenstructure mismatch.

\begin{table}[t!]
\centering
\caption{Similarity metric $\eta$ under positioning ($\sigma_g=0$) and gain ($\sigma_p=0$) perturbations in the \emph{urban} scenario.}
\renewcommand{\arraystretch}{1.5}
\setlength{\tabcolsep}{4pt}
\scriptsize
\begin{tabular}{c|ccc|ccc}
\hline
& \multicolumn{3}{c|}{\textbf{Positioning ($\sigma_p$ [m])}}
& \multicolumn{3}{c}{\textbf{Gain ($\sigma_g$)}} \\
\textbf{Config.}
& 0.1 & 1 & 10
& 0.1 & 0.25 & 0.5 \\
\hline
$4,32$
& 1.000 & 1.000 & 0.976
& 0.995 & 0.970 & 0.918 \\
$8,64$
& 1.000 & 0.995 & 0.934
& 0.995 & 0.965 & 0.917 \\
$16,128$
& 1.000 & 0.991 & 0.882
& 0.995 & 0.960 & 0.905 \\
\hline
\end{tabular}
\label{tab:similarity_combined}
\end{table}

Table~\ref{tab:similarity_combined} reports the similarity metric under both positioning and gain perturbations for different system configurations. Under positioning errors ($\sigma_g=0$), as discussed in Sec.~\ref{sec:DT_perturbations_metrics_nmse}, perturbations jointly affect AoA, AoD, and delay, resulting in a distortion of the channel eigenstructure. For centimeter-level perturbations, the similarity remains close to one, indicating that the eigenstructure is accurately preserved. However, as $\sigma_p$ increases, a clear degradation is observed, which becomes more pronounced for larger system dimensions. This behavior reflects the higher sensitivity of high-resolution systems to geometric mismatches, as the finer ST resolution makes the eigenstructure more susceptible to perturbations.

In contrast, under gain perturbations ($\sigma_p=0$), $\eta$ mainly reflects the distortion of the eigenvalue profile, while the channel subspace remains largely unchanged. The perturbation primarily redistributes power across the propagation modes, with limited impact on the dominant eigenstructure. As a result, $\eta$ remains high across all configurations, even for relatively large values of $\sigma_g$, and exhibits only a weak dependence on the system dimension.


\subsection{NMSE Analysis under Perturbations}
\label{sec:nmse_analysis}

\begin{figure}[t!]
\centering
\includegraphics[width=0.89\linewidth]{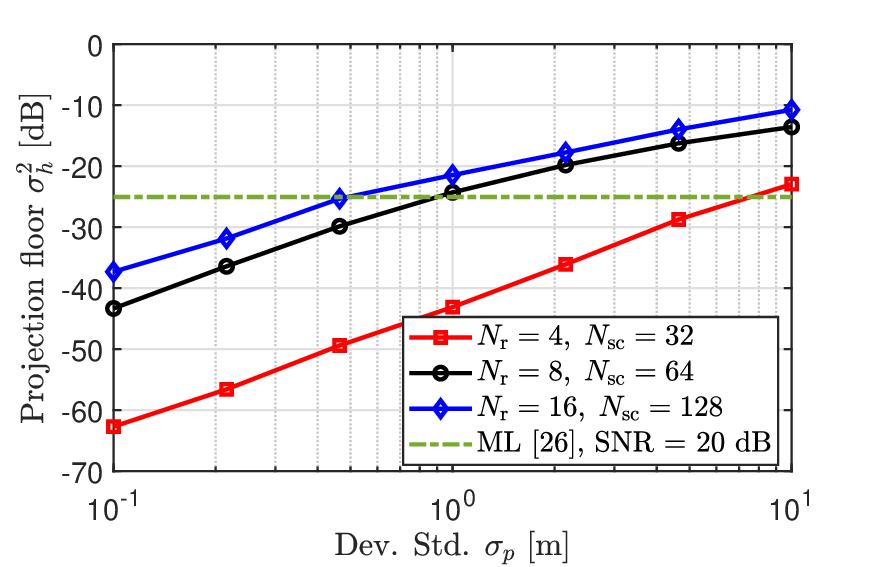}
\caption{Projection error floor versus positioning error standard deviation $\sigma_p$ for different system configurations in the \emph{urban} scenario}.
\label{fig:projection_floor}
\end{figure}

We next consider the projection error term introduced in~\eqref{eq:Channel_projection_LR}, which captures the residual channel energy outside the DT-provided subspace. In the high SNR regime ($\sigma_w^2 \rightarrow 0$), this term corresponds to the projection error, which determines the NMSE of both LR and MMSE estimators, and thus defines a fundamental performance floor. In particular, this analysis provides a practical indication of the maximum tolerable positioning error for a given system configuration when compared to baseline approaches (where the ML estimator is taken as reference). Fig.~\ref{fig:projection_floor} reports this projection error floor as a function of the positioning error standard deviation. As expected, the floor increases with $\sigma_p$. For small positioning errors (roughly $\sigma_p < 0.2$ m), the floor remains very low (below approximately $-30$ dB), indicating that the DT accurately reconstructs the eigenstructure of the channel. However, for $\sigma_p \approx 1$ m, the floor increases, revealing a substantial loss due to eigenstructure mismatches especially for higher-resolution configurations. More specifically, in the configuration with $N_{\mathrm r}=16$ and $N_{\mathrm sc}=128$, the admissible positioning error lies approximately in the range $0.4$--$0.6$ m when compared to the ML estimator at $\mathrm{SNR}=20$ dB, while with $N_{\mathrm r}=8$ and $N_{\mathrm sc}=64$ the maximum tolerable positioning error is roughly $1~\mathrm{m}$.

\begin{figure}[t!]
\centering
\includegraphics[width=0.89\linewidth]{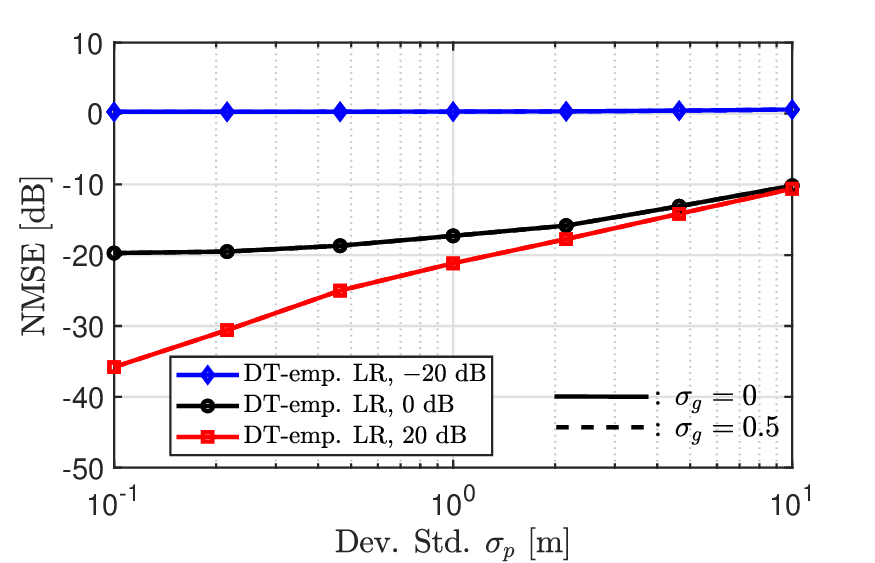}
\caption{NMSE of the DT-empowered LR estimator versus positioning standard deviation $\sigma_p$ in the \emph{urban} scenario, for different SNRs and gain perturbations levels. Solid and dashed curves correspond to $\sigma_g=0$ and $\sigma_g=0.5$, respectively.}
\label{Fig:gain_impact_LR}
\end{figure}

We then evaluate the NMSE performance of the DT-empowered LR estimator under joint gain and positioning perturbations for finite SNR values and for a system configuration with $N_{\mathrm r}=16$ and $N_{\mathrm sc}=128$. To this end, Fig.~\ref{Fig:gain_impact_LR} shows the NMSE performance as a function of $\sigma_p$, for different SNR values and gain perturbation levels in the \emph{urban} scenario. The DT-empowered LR estimator is essentially insensitive to gain perturbations, as the curves corresponding to $\sigma_g = 0$ and $\sigma_g = 0.5$ overlap across the entire $\sigma_p$ range for all SNR values. This behavior provides a direct validation of the theoretical analysis, confirming that LR estimation depends only on the channel subspace, which remains unaffected by gain perturbations as long as the dominant propagation structure is preserved.

\begin{figure*}[t!]
\centering

\begin{subfigure}[b]{0.32\textwidth}
\centering
\includegraphics[width=\linewidth]{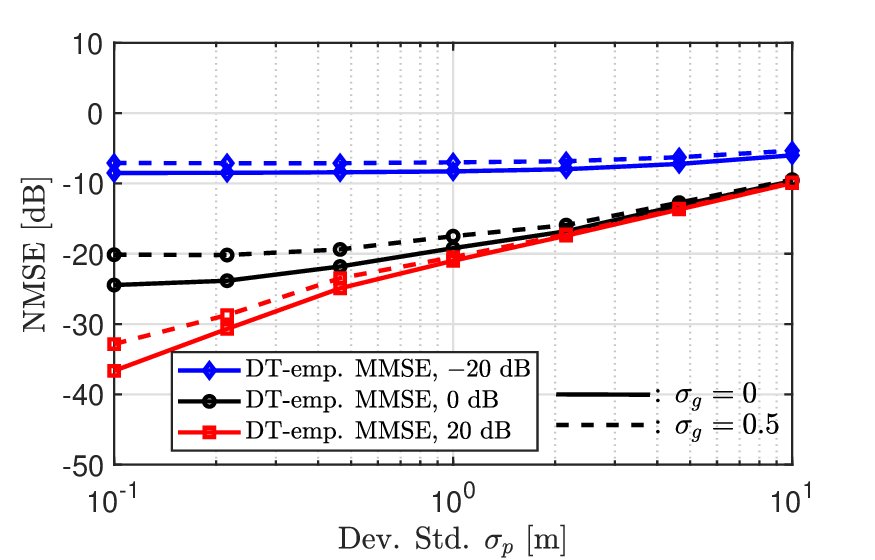}
\caption{\emph{Urban}}
\label{fig:MMSE_gain_urban}
\end{subfigure}
\hfill
\begin{subfigure}[b]{0.32\textwidth}
\centering
\includegraphics[width=\linewidth]{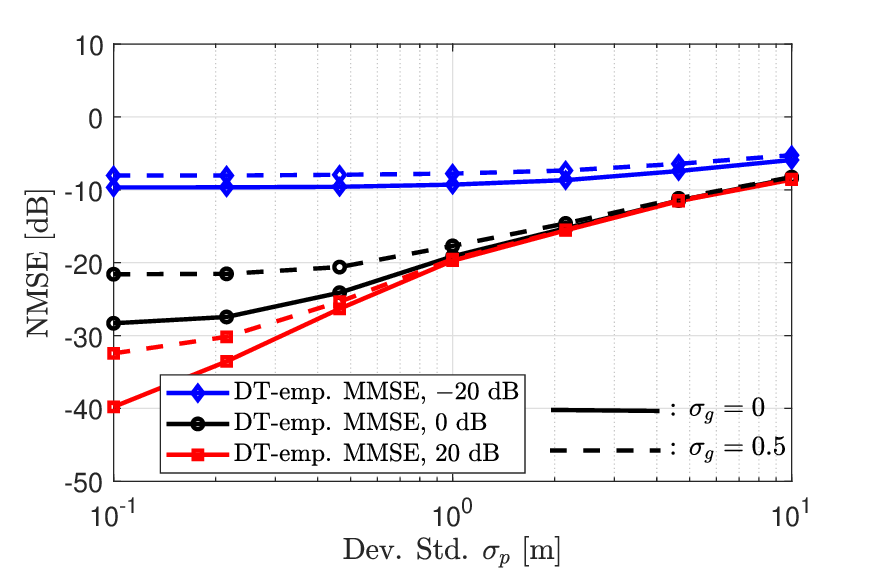}
\caption{\emph{Suburban}}
\label{fig:MMSE_gain_suburban}
\end{subfigure}
\hfill
\begin{subfigure}[b]{0.32\textwidth}
\centering
\includegraphics[width=\linewidth]{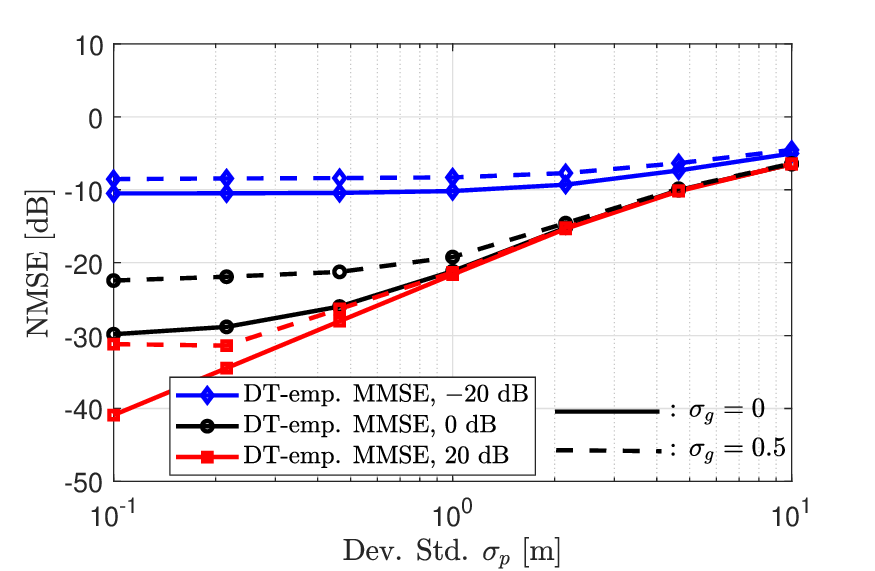}
\caption{\emph{Rural}}
\label{fig:MMSE_gain_rural}
\end{subfigure}

\caption{NMSE of the DT-empowered MMSE estimator versus positioning standard deviation $\sigma_p$ for different SNR values, gain perturbations levels, and scenarios. Solid and dashed curves correspond to $\sigma_g=0$ and $\sigma_g=0.5$, respectively.}
\label{Fig:gain_impact_MMSE_scenarios}
\end{figure*}

We further consider the DT-empowered MMSE estimator in the three propagation scenarios. Fig.~\ref{Fig:gain_impact_MMSE_scenarios} reports the NMSE performance as a function of $\sigma_p$ in the \emph{urban}, \emph{suburban}, and \emph{rural} scenarios. In contrast to LR estimation, MMSE exhibits a moderate sensitivity to gain perturbations because it depends not only on the dominant subspace but also on the covariance eigenvalue profile, which is directly affected by gain errors through the eigenvalue-dependent shrinkage coefficients. Nevertheless, the resulting degradation remains limited compared to that induced by positioning errors.

Although the absolute NMSE slightly differs across the three environments in the absence of perturbations because of the different propagation characteristics and channel sparsity, the impact of positioning and gain errors follows the same qualitative behavior. Gain perturbations only moderately affect the MMSE estimator, yielding less than $2$ dB degradation at $\sigma_p=1~\mathrm{m}$ across all considered SNR values. In contrast, positioning errors remain the dominant impairment, leading to nearly identical degradation trends in all considered propagation environments. These results confirm that the observed behavior is consistent across environments with substantially different propagation characteristics, thereby supporting the generality of the previous analysis.

More importantly, these results consistently show that positioning errors represent the primary source of degradation. As $\sigma_p$ increases, the NMSE deteriorates across all SNR regimes, reflecting the impact of eigenstructure mismatch induced by geometric perturbations. This effect is particularly pronounced at high SNR, where the noise contribution becomes negligible and the performance is dominated by subspace mismatch. At low SNR, the NMSE is dominated by noise and remains nearly constant across different values of $\sigma_p$. As the SNR increases, the noise contribution diminishes and the estimation performance becomes limited by eigenstructure mismatch, leading to an NMSE floor determined by the projection error. This transition clearly highlights the fundamental role of subspace mismatch in limiting the performance of DT-empowered estimators and confirms the key theoretical insight that positioning errors, rather than gain perturbations, constitute the main bottleneck in DT-empowered channel estimation.

\subsection{Performance Evaluation Against Baseline Approaches}
\label{sec:performance_evaluation}

\begin{figure*}[t!]
\centering

\begin{subfigure}[b]{0.32\textwidth}
\centering
\includegraphics[width=\linewidth]{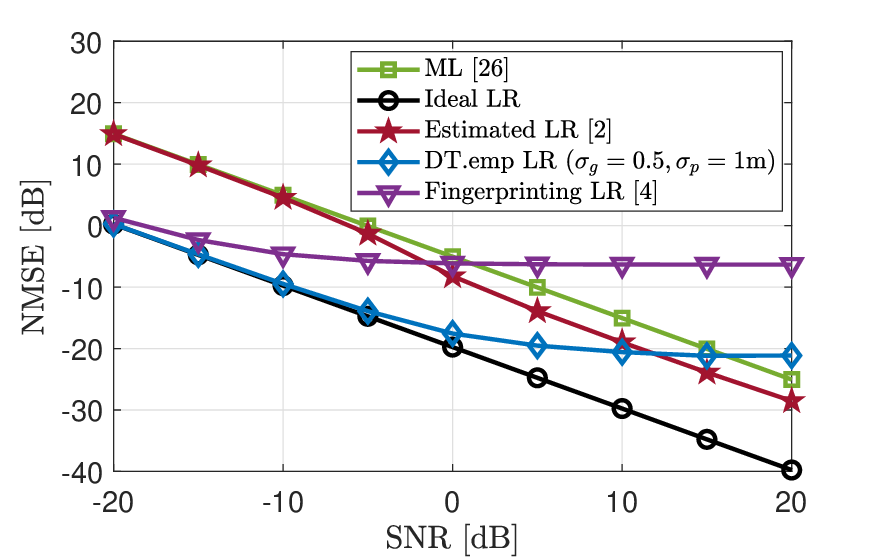}
\caption{\emph{Urban}}
\end{subfigure}
\hfill
\begin{subfigure}[b]{0.32\textwidth}
\centering
\includegraphics[width=\linewidth]{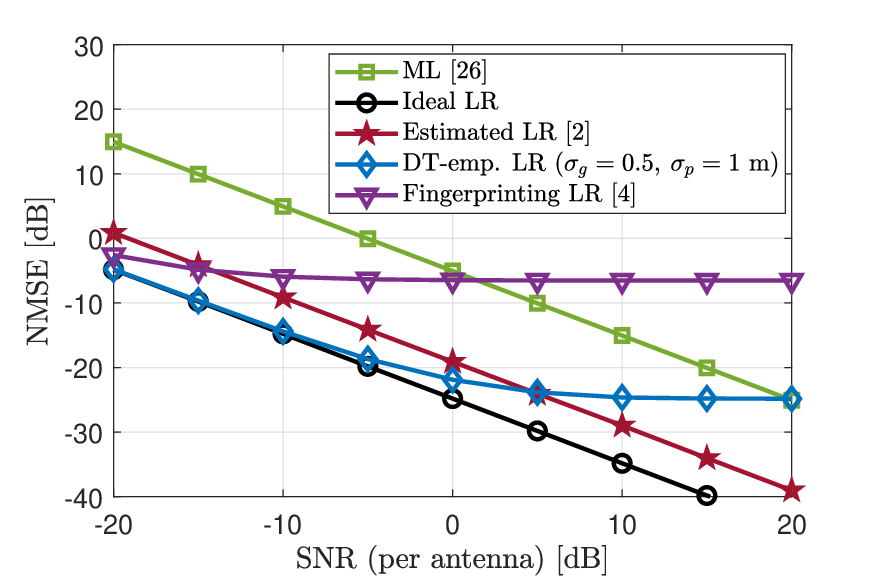}
\caption{\emph{Suburban}}
\end{subfigure}
\hfill
\begin{subfigure}[b]{0.32\textwidth}
\centering
\includegraphics[width=\linewidth]{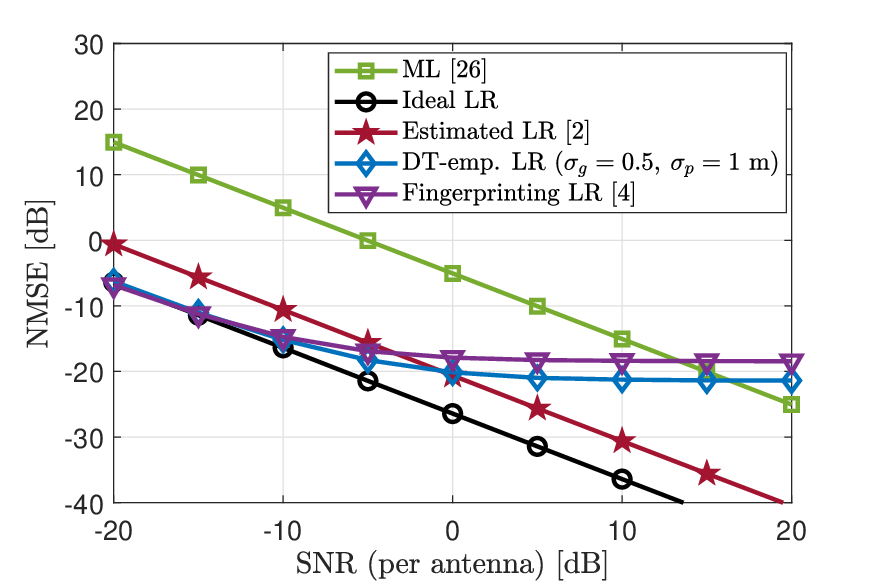}
\caption{\emph{Rural}}
\end{subfigure}

\vspace{2mm}

\begin{subfigure}[b]{0.32\textwidth}
\centering
\includegraphics[width=\linewidth]{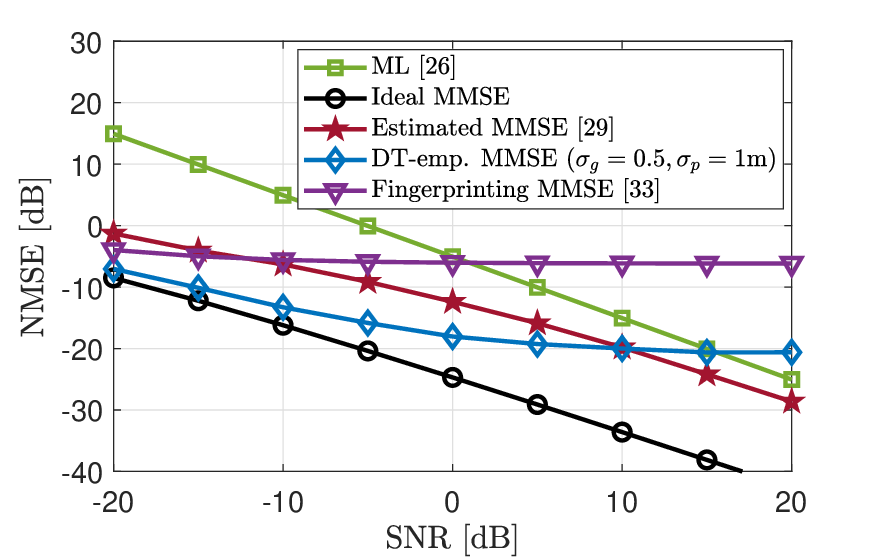}
\caption{\emph{Urban}}
\end{subfigure}
\hfill
\begin{subfigure}[b]{0.32\textwidth}
\centering
\includegraphics[width=\linewidth]{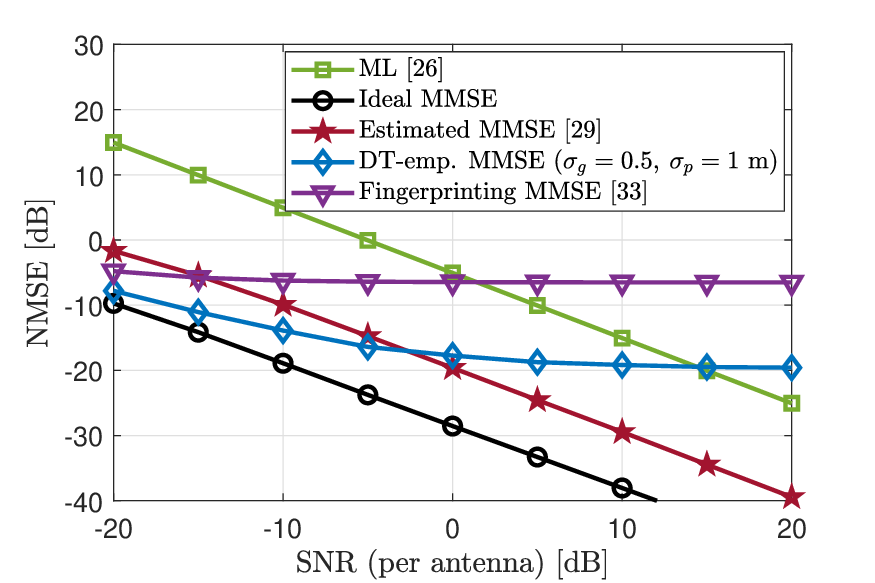}
\caption{\emph{Suburban}}
\end{subfigure}
\hfill
\begin{subfigure}[b]{0.32\textwidth}
\centering
\includegraphics[width=\linewidth]{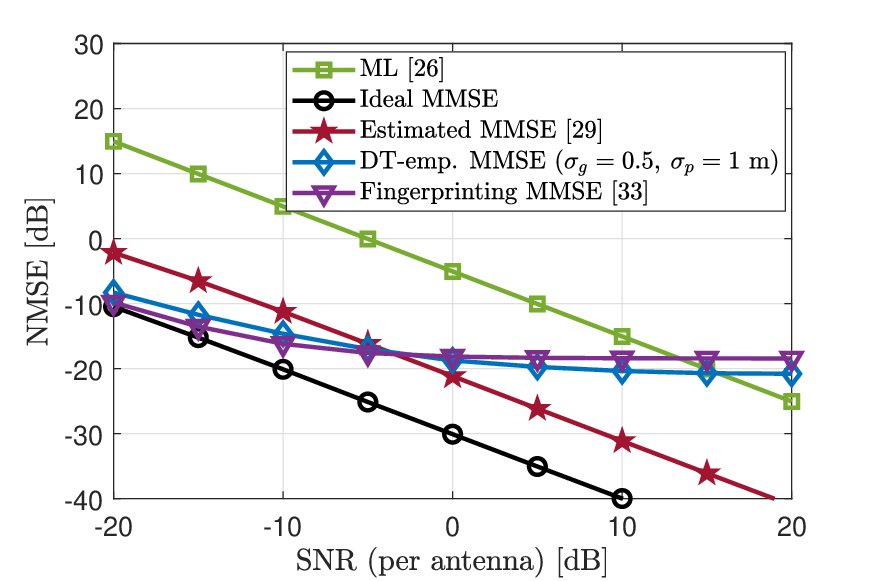}
\caption{\emph{Rural}}
\end{subfigure}

\caption{NMSE versus SNR for the DT-empowered LR and MMSE estimators and baseline methods under $\sigma_p=1~\mathrm{m}$ and $\sigma_g=0.5$ across the considered scenarios.}
\label{Fig:NMSE_comp}
\end{figure*}

Finally, Fig.~\ref{Fig:NMSE_comp} compares the proposed DT-empowered estimators with baseline approaches as a function of the SNR in the \emph{urban}, \emph{suburban}, and \emph{rural} scenarios. Fig.~\ref{Fig:NMSE_comp}(a), Fig.~\ref{Fig:NMSE_comp}(b), and Fig.~\ref{Fig:NMSE_comp}(c) report the LR results in the \emph{urban}, \emph{suburban}, and \emph{rural} scenarios, respectively, whereas Fig.~\ref{Fig:NMSE_comp}(d), Fig.~\ref{Fig:NMSE_comp}(e), and Fig.~\ref{Fig:NMSE_comp}(f) report the corresponding MMSE results.

For LR estimation, Fig.~\ref{Fig:NMSE_comp}(a), Fig.~\ref{Fig:NMSE_comp}(b), and Fig.~\ref{Fig:NMSE_comp}(c) show that the DT-empowered estimator provides a clear advantage in the low- and medium-SNR regimes across the three scenarios, owing to the more accurate subspace information provided by the DT. As the SNR increases, the performance gradually saturates because of the residual mismatch induced by positioning errors. Under the considered perturbation setting, $\sigma_p=1~\mathrm{m}$ and $\sigma_g=0.5$, this saturation is consistent with the projection error floor observed in Fig.~\ref{fig:projection_floor}. The absolute NMSE values vary across scenarios because the \emph{suburban} and \emph{rural} channels generally exhibit a more compact dominant ST subspace than the \emph{urban} channel, due to their lower multipath richness.

For MMSE estimation, Fig.~\ref{Fig:NMSE_comp}(d), Fig.~\ref{Fig:NMSE_comp}(e), and Fig.~\ref{Fig:NMSE_comp}(f) show a similar behavior. The DT-empowered MMSE estimator performs close to the ideal MMSE estimator in the low-SNR regime and provides significant gains over the ML estimator and the fingerprinting-based approach. At higher SNR, the performance is limited by the residual mismatch between the true and DT-reconstructed covariance. Since MMSE depends not only on the dominant subspace but also on the covariance eigenvalue profile, it is more sensitive than LR to gain perturbations. Nevertheless, consistently with Fig.~\ref{Fig:gain_impact_MMSE_scenarios}, the dominant limitation remains the positioning-induced eigenspace mismatch.

The comparison across scenarios also clarifies the behavior of the fingerprinting-based benchmark. In the \emph{rural} scenario, this baseline becomes closer to the DT-empowered estimators because the propagation is sparser and the spatial channel variations are smoother. In this case, the nearest-grid covariance is more representative of the actual UE channel, reducing the mismatch introduced by grid discretization. Fingerprinting-based approaches are, hence, inherently limited by the spatial resolution of the underlying grid. The discretization introduces mismatches in the reconstructed propagation structure, leading to artificial shadowing effects (i.e., spurious path birth/loss). This degrades the estimated channel eigenstructure, explaining the observed performance gap and saturation behavior.

Overall, these results show that the DT-empowered framework provides a reliable and physically consistent prior for channel estimation across outdoor scenarios with different building densities, propagation sparsity, and multipath richness. Performance gains are most significant at low SNR and remain robust to gain perturbations, while ultimately limited by UE positioning accuracy.

\section{Conclusion}
\label{sec:conclusion}
This paper investigated the robustness of DT-empowered channel estimation in wideband MIMO systems, where site-specific propagation features extracted via RT are used to reconstruct the channel covariance eigenstructure and enable LR and MMSE estimation. By developing first-order perturbation models, we showed that positioning errors induce geometric mismatches that distort the channel subspace, while EM calibration errors mainly affect the eigenvalue profile. This structural distinction explains the different sensitivity of the estimators, with LR being inherently robust to EM perturbations and MMSE affected by both subspace and eigenvalue mismatches. Numerical results across urban, suburban, and rural scenarios confirmed that the proposed framework exhibits consistent behavior across environments with different building densities, propagation sparsity, and multipath richness. In all considered scenarios, positioning inaccuracies are the dominant source of performance degradation, whereas EM calibration errors have a comparatively limited impact. Overall, the results highlight the critical role of accurate geometric information in DT-enabled communications. Future work will focus on extending the analysis to large mismatches and imperfect 3D scene reconstruction, as well as on integrating adaptive techniques to mitigate subspace distortions in practical DT systems. A further relevant direction is the measurement-based validation of the proposed framework: controlled channel measurements at known UE positions would allow the RT-reconstructed covariance eigenstructure to be compared with the sample covariance and would enable direct quantification of the perturbation levels induced by realistic positioning and EM calibration errors.

\appendices
\section{First-Order Approximation of ST features}
\label{Appendix:B}

In this appendix, we derive first-order Taylor approximations of distance, azimuth, and elevation angles under small perturbations of a 3D displacement vector.  used to characterize geometric uncertainties in ST features. Consider a real-valued function 
$g : \mathbb{R}^3 \rightarrow \mathbb{R}$, 
with argument $\mathbf{d} = [d_1,d_2,d_3]^{\mathsf T}\in \mathbb{R}^3$. 
Let $\overline{\mathbf{d}} = [\overline{d}_1,\overline{d}_2,\overline{d}_3]^{\mathsf T}\in \mathbb{R}^3$ 
be an arbitrary point, and let 
$\boldsymbol{\epsilon} \in \mathbb{R}^3$ 
denote a small perturbation such that 
$\mathbf{d} = \overline{\mathbf{d}} + \boldsymbol{\epsilon}$. The first-order approximation of $g$ around 
$\overline{\mathbf{d}}$ is given by
\begin{equation}
g(\overline{\mathbf{d}}+\boldsymbol{\epsilon})
=
g(\overline{\mathbf{d}})
+
\nabla g(\overline{\mathbf{d}})^{\mathsf T}\boldsymbol{\epsilon}
+
o(\|\boldsymbol{\epsilon}\|),
\end{equation}
where $\nabla g(\overline{\mathbf{d}}) \in \mathbb{R}^3$ 
denotes the gradient of $g$ evaluated in $\overline{\mathbf{d}}$.

In the specific case of $g(\mathbf{d})=\|\mathbf{d}\|$, 
the gradient evaluated in 
$\overline{\mathbf{d}}$, with $\overline{\mathbf{d}}\neq \mathbf{0}$, 
\begin{equation}
\frac{\partial g(\overline{\mathbf{d}})}{\partial d_i}
=\frac{\overline{d}_i}{\overline{\|\mathbf{d}\|}},\quad \text{for}\: i=1,2,3,
\end{equation}
is $
\nabla g(\overline{\mathbf{d}})
= \overline{\mathbf{d}}/\|\overline{\mathbf{d}}\| = \mathbf{a} $. This yields the first-order approximation
\begin{equation}
g(\overline{\mathbf{d}}+\boldsymbol{\epsilon})
\approx
g(\overline{\mathbf{d}})
+
\mathbf{a}^{\mathsf T}\boldsymbol{\epsilon}.
\end{equation}

For the particular case of $g(\mathbf{d})=\mathrm{atan2}(d_2,d_1)$,
assume that $\overline{\mathbf d}$ belongs to the open set
\[
\mathcal{D}
=
\mathbb{R}^2
\setminus
\left(
\{(0,0)\}
\cup
\{(d_1,d_2)\in\mathbb{R}^2 : d_1<0,\ d_2=0\}
\right),
\]
where $\mathrm{atan2}$ is differentiable. The gradient evaluated in 
$\overline{\mathbf{d}}$ is 
\begin{equation}
\nabla g(\overline{\mathbf{d}})
=
\frac{1}{\rho^2}
\begin{bmatrix}
-\overline{d}_2 \\
\overline{d}_1 \\
0
\end{bmatrix}
=
\frac{1}{\rho^2}\mathbf{a},
\end{equation}
where $\rho=\sqrt{\overline{d}_1^2+\overline{d}_2^2}$ and 
$\mathbf{a}=
\begin{bmatrix}
-\overline{d}_2 &
\overline{d}_1 &
0
\end{bmatrix}^{\mathsf T}$. The first-order approximation holds locally as
\begin{equation}
g(\overline{\mathbf{d}}+\boldsymbol{\epsilon})
\approx
g(\overline{\mathbf{d}})
+
\frac{1}{\rho^2}\mathbf{a}^{\mathsf T}\boldsymbol{\epsilon}.
\end{equation}

In the specific case of 
$g(\mathbf{d})=\arcsin\!\left(d_3/\|\mathbf{d}\|\right)$,
the gradient evaluated in 
$\overline{\mathbf{d}}$, with 
$\overline{\mathbf{d}}\neq\mathbf{0}$ and 
$\overline{d}_1^2+\overline{d}_2^2\neq 0$ is 
\begin{equation}
\nabla g(\overline{\mathbf{d}})
=
\frac{1}{R^2\rho}
\begin{bmatrix}
-\overline{d}_3\overline{d}_1 \\
-\overline{d}_3\overline{d}_2 \\
\overline{d}_1^2+\overline{d}_2^2
\end{bmatrix}
=
\frac{1}{R^2\rho}\mathbf{c},
\end{equation}
where $R^2=\|\overline{\mathbf{d}}\|^2$ and
$\mathbf{c}=
\begin{bmatrix}
-\overline{d}_3\overline{d}_1 &
-\overline{d}_3\overline{d}_2 &
\overline{d}_1^2+\overline{d}_2^2
\end{bmatrix}^{\mathsf T}$.
This yields the first-order approximation
\begin{equation}
g(\overline{\mathbf{d}}+\boldsymbol{\epsilon})
\approx
g(\overline{\mathbf{d}})
+
\frac{1}{R^2\rho}\mathbf{c}^{\mathsf T}\boldsymbol{\epsilon}.
\end{equation}

\section{Material-Induced Perturbations}
\label{Appendix:C}

In this appendix, we derive the sensitivity factors for material-induced perturbations and their first-order impact on the path gain.

\subsection{Sensitivity Factors for the Reflection Coefficient}

Consider the reflection coefficient expressed as a function 
\(
\Gamma(\varepsilon,\theta)
\)
of the complex relative permittivity 
\(
\varepsilon \in \mathbb{C}
\)
and the incidence angle 
\(
\theta \in [0,\pi/2].
\)
The sensitivity factor is defined as
\begin{equation}
c_{p,k}
=
\left.
\frac{\partial \Gamma(\varepsilon,\theta_{p,k})}
{\partial \varepsilon}
\right|_{\varepsilon=\varepsilon^r_{p,k}}.
\end{equation}

For TE polarization, the reflection coefficient is given by
\begin{equation}
\Gamma^{\mathrm{TE}}(\varepsilon,\theta)
=
\frac{\cos\theta
-
\sqrt{\varepsilon-\sin^2\theta}}
{\cos\theta
+
\sqrt{\varepsilon-\sin^2\theta}}.
\end{equation}
Let
\(
s(\varepsilon)
=
\sqrt{\varepsilon-\sin^2\theta}
\), with $
\frac{\partial s(\varepsilon)}{\partial \varepsilon}=\frac{1}{2s(\varepsilon)}$. The derivative $\partial \Gamma^{\mathrm{TE}}/\partial \varepsilon$ is
\begin{equation}
\begin{split}
\frac{\partial \Gamma^{\mathrm{TE}}}{\partial \varepsilon}&=
-\frac{
2\cos\theta
}{
(\cos\theta + s(\varepsilon))^2
}
\frac{\partial s(\varepsilon)}{\partial \varepsilon}.
\end{split}
\end{equation}
This results in a sensitivity factor for TE polarization
\begin{equation}
c^{\mathrm{TE}}_{p,k}
=
-\frac{\cos\theta_{p,k}}
{\sqrt{\varepsilon^r_{p,k}-\sin^2\theta_{p,k}}
\left(
\cos\theta_{p,k}
+
\sqrt{\varepsilon^r_{p,k}-\sin^2\theta_{p,k}}
\right)^2}.
\end{equation}

For TM polarization, the reflection coefficient is given by
\begin{equation}
\Gamma^{\mathrm{TM}}(\varepsilon,\theta)
=
\frac{\varepsilon\cos\theta
-
\sqrt{\varepsilon-\sin^2\theta}}
{\varepsilon\cos\theta
+
\sqrt{\varepsilon-\sin^2\theta}}.
\end{equation}
The derivative $\partial \Gamma^{\mathrm{TM}}/\partial \varepsilon$ is
\begin{equation}
\frac{\partial \Gamma^{\mathrm{TM}}}{\partial \varepsilon}
=
\frac{
2\cos\theta\, s(\varepsilon)
-
2\varepsilon\cos\theta\,\frac{\partial s(\varepsilon)}{\partial \varepsilon}
}{
(\varepsilon\cos\theta+s(\varepsilon))^2
}.
\end{equation}
This yields the sensitivity factor
\begin{equation}
c^{\mathrm{TM}}_{p,k}
=
\frac{\cos\theta_{p,k}
\left(
\sqrt{\varepsilon^r_{p,k}-\sin^2\theta_{p,k}}
-
\frac{\varepsilon^r_{p,k}}
{\sqrt{\varepsilon^r_{p,k}-\sin^2\theta_{p,k}}}
\right)}
{\left(
\varepsilon^r_{p,k}\cos\theta_{p,k}
+
\sqrt{\varepsilon^r_{p,k}-\sin^2\theta_{p,k}}
\right)^2}.
\end{equation}

For general polarization conditions, the effective reflection coefficient can be modeled as a combination of TE and TM components. This representation should be interpreted as an effective parametrization rather than a strict physical decomposition.

\subsection{First-Order Propagation to the Path Gain}

Consider the DT path gain
\begin{equation}
\hat{\Omega}_p
=
\Omega_p^{(0)}
\prod_{k=1}^{K_p}
|\hat{\Gamma}_{p,k}|^2.
\end{equation}
Using the first-order approximation $\hat{\Gamma}_{p,k}
\approx
\Gamma_{p,k}
+
c_{p,k}\,\delta\varepsilon^r_{p,k}$, we expand the squared magnitude as
\begin{equation}
|\hat{\Gamma}_{p,k}|^2
=
|\Gamma_{p,k}|^2
+
2\,\Re\{\Gamma_{p,k}^* c_{p,k}\delta\varepsilon^r_{p,k}\}
+
|c_{p,k}|^2|\delta\varepsilon^r_{p,k}|^2.
\end{equation}
Under the small-perturbation assumption, the quadratic term $|c_{p,k}|^2|\delta\varepsilon^r_{p,k}|^2$ is neglected. Hence,
\begin{equation}
|\hat{\Gamma}_{p,k}|^2
\approx
|\Gamma_{p,k}|^2
+
2\,\Re\{\Gamma_{p,k}^* c_{p,k}\delta\varepsilon^r_{p,k}\}.
\end{equation}
Substituting into the product and retaining only first-order terms yields
\begin{equation}
\begin{split}
\hat{\Omega}_p
\approx
&\Omega_p^{(0)}
\prod_{k=1}^{K_p}
|\Gamma_{p,k}|^2
\\
&+
\Omega_p^{(0)}
\sum_{k=1}^{K_p}
\left(
2\,\Re\{\Gamma_{p,k}^* c_{p,k}\delta\varepsilon^r_{p,k}\}
\prod_{\substack{m=1 \\ m\neq k}}^{K_p}
|\Gamma_{p,m}|^2
\right).
\end{split}
\end{equation}
Recognizing that $\Omega_p
=
\Omega_p^{(0)}
\prod_{m=1}^{K_p}
|\Gamma_{p,m}|^2$, it follows that $\Omega_p^{(0)}\prod_{m\neq k}|\Gamma_{p,m}|^2=\Omega_p/|\Gamma_{p,k}|^2$, and hence
\begin{equation}
\hat{\Omega}_p
\approx
\Omega_p
+
\sum_{k=1}^{K_p}
2\,\Omega_p\,
\Re\!\left\{
\frac{\Gamma_{p,k}^*}{|\Gamma_{p,k}|^2}\,
c_{p,k}\delta\varepsilon^r_{p,k}
\right\}.
\end{equation}
Since $\Gamma_{p,k}\Gamma_{p,k}^*=|\Gamma_{p,k}|^2$, we have $\Gamma_{p,k}^*/|\Gamma_{p,k}|^2=1/\Gamma_{p,k}$, which yields
\begin{equation}
\hat{\Omega}_p
\approx
\Omega_p
+
\Omega_p\sum_{k=1}^{K_p}
\Re\!\left\{
\frac{2c_{p,k}}{\Gamma_{p,k}}\,
\delta\varepsilon^r_{p,k}
\right\}.
\end{equation}
Defining the complex sensitivity coefficient $
\alpha_{p,k}
=
2c_{p,k}/\Gamma_{p,k}$, the first-order perturbation model can be compactly expressed as
\begin{equation}
\hat{\Omega}_p
\approx
\Omega_p
+
\Omega_p\sum_{k=1}^{K_p}
\Re\!\left\{
\alpha_{p,k}\,
\delta\varepsilon^r_{p,k}
\right\}.
\end{equation}

\bibliographystyle{IEEEtran}
\bibliography{bibliography}

\end{document}